\documentclass{article}

\usepackage{tabularx,booktabs}

\usepackage{PRIMEarxiv}

\usepackage[utf8]{inputenc} 
\usepackage[T1]{fontenc}    
\usepackage{hyperref}       
\usepackage{url}            
\usepackage{booktabs}       
\usepackage{amsfonts}       
\usepackage{nicefrac}       
\usepackage{microtype}      
\usepackage{lipsum}
\usepackage{fancyhdr}       
\usepackage{graphicx}       
\graphicspath{{media/}}     
\usepackage{seqsplit}
\usepackage{geometry}

\usepackage{tabularx,booktabs}
\usepackage{wrapfig}
\usepackage[utf8]{inputenc} 
\usepackage[T1]{fontenc}    
\usepackage{hyperref}       
\usepackage{url}            
\usepackage{booktabs}       
\usepackage{amsfonts}       
\usepackage{nicefrac}       
\usepackage{microtype}      
\usepackage{lipsum}
\usepackage{fancyhdr}       
\usepackage{graphicx}       
\graphicspath{{media/}}     
\usepackage{seqsplit}
\usepackage{geometry}

\usepackage{subcaption}
\usepackage{amsthm,amssymb,amsfonts}
\usepackage{tikz,lipsum,lmodern}
\usepackage[most]{tcolorbox}
\definecolor{cream}{RGB}{222,217,201}

\newtcolorbox{Box1}[2][]{
                lower separated=false,
                colback=white!80!gray,
colframe=black, fonttitle=\bfseries,
colbacktitle=black!50!gray,
coltitle=black,
enhanced,
attach boxed title to top left={xshift=0.5cm,yshift=-2mm},
title=#2,
boxrule=0.5pt,
boxed title style={colframe=black, boxrule=0.5pt},
#1}

\title{{Rapid and Automated Alloy Design with Graph Neural Network-Powered LLM-Driven Multi-Agent Systems}

\thanks{\textit{\underline{Citation}}: 
\textbf{A. Ghafarollahi, M.J. Buehler. arXiv, DOI:000000/11111., 2024}} 
}

\author{
  Alireza Ghafarollahi \\
  Laboratory for Atomistic and Molecular Mechanics (LAMM)\\Massachusetts Institute of Technology\\ 77 Massachusetts Ave.\\ Cambridge, MA 02139, USA 
   \And
  Markus J. Buehler \\
  Laboratory for Atomistic and Molecular Mechanics (LAMM)  \\
  Center for Computational Science and Engineering\\ Schwarzman College of Computing\\ Massachusetts Institute of Technology\\77 Massachusetts Ave.\\Cambridge, MA 02139, USA\\ \\
  Correspondence: \texttt{mbuehler@MIT.EDU} \\
}

\begin{document}
\maketitle

\begin{abstract}
A multi-agent AI model is used to automate the discovery of new metallic alloys, integrating multimodal data and external knowledge including insights from physics via atomistic simulations. Our multi-agent system features three key components: (a) a suite of LLMs responsible for tasks such as reasoning and planning, (b) a group of AI agents with distinct roles and expertise that dynamically collaborate, and (c) a newly developed graph neural network (GNN) model for rapid retrieval of key physical properties. A set of LLM-driven AI agents collaborate to automate the exploration of the vast design space of MPEAs, guided by predictions from the GNN. We focus on the NbMoTa family of body-centered cubic (bcc) alloys, modeled using an ML-based interatomic potential, and target two key properties: the Peierls barrier and solute/screw dislocation interaction energy. Our GNN model accurately predicts these atomic-scale properties, providing a faster alternative to costly brute-force calculations and reducing the computational burden on multi-agent systems for physics retrieval. This AI system revolutionizes materials discovery by reducing reliance on human expertise and overcoming the limitations of direct all-atom simulations. By synergizing the predictive power of GNNs with the dynamic collaboration of LLM-based agents, the system autonomously navigates vast alloy design spaces, identifying trends in atomic-scale material properties and predicting macro-scale mechanical strength, as demonstrated by several computational experiments. This approach accelerates the discovery of advanced alloys and holds promise for broader applications in other complex systems, marking a significant step forward in automated materials design.
\end{abstract}

\keywords{Multi-agent system, Large language model \and Deep learning \and Graph neural networks \and complex alloys \and Materials design \and Scientific machine learning}

\section{Introduction}
Multi-principal element alloys (MPEAs) represent a relatively new and innovative class of materials comprising three or more elements and showing remarkable mechanical properties superior to their pure or dilute counterparts, such as mechanical strength, fracture toughness, ductility, and resistance to hydrogen embrittlement. \cite{ye2016high, miracle2017critical, george2019high, george2020high}. The BCC refractiry MPEAs of the family Cr-Mo-W-Nb-V-Ta-Ti-Zr-Hf have recently gained particular interest due to their exceptional high-temperaute strength retention, exceeding the capabilities of current superalloys. \cite{senkov2010refractory, senkov2011mechanical, senkov2011microstructure, senkov2018development, senkov2018compositional, han2018microstructures, senkov2019high, xiong2023refractory} With millions of possible compositions within just one single crystal, these complex systems offer significant potential to achieve tailored properties for specialized applications. However, navigating the immense multicomponent HEA compositional space to guide alloy discovery with optimized or desired properties poses significant challenges. Developing mechanistic multi-scale theories that connect atomisitc-level phenomenon to microscopic material properties, such as temperature dependent yield sterss, have been instrumental in exploring this design space. \cite{varvenne2016theory, varvenne2017solute, rao2019solution, ghafarollahi2020theory, ghafarollahi2021theory, ghamaresca2020mechanistic, rao2021theory, kubilay2021high, ghafarollahi2022screw, baruffi2022screw, rao2022theory}

While theoretical models hold significant potential for predicting the macroscopic properties of MPEAs and navigating their vast design space, a key limitation lies in the computational cost of obtaining the necessary input parameters, which typically rely on atomistic simulations. This challenge is particularly acute for body-centered cubic (BCC) materials, where plasticity is controlled by the motion of screw dislocations \cite{maresca2020theory, ghafarollahi2022screw}. In contrast to edge dislocation theories, which offer a simplified approach by using misfit volume as a primary parameter \cite{varvenne2016theory, varvenne2017solute, maresca2020mechanistic}, no such simplification exists for screw dislocation theories. As a result, these models require the computation of critical parameters via expensive atomistic simulations. Two key quantities in screw dislocation theories are the Peierls barrier and the solute-screw interaction energy \cite{maresca2020theory, ghafarollahi2022screw}. The Peierls barrier represents the intrinsic lattice resistance to dislocation motion, while the solute-screw interaction energy describes the influence of solute atoms on the dislocation's behavior. Both factors serve as energy barriers to dislocation motion, and their calculation is further complicated by random solute fluctuations around the dislocation. Atomistic simulations are commonly used to determine these parameters in alloys \cite{ghafarollahi2019solute}, but applying them to multi-component alloys presents two significant challenges: (a) the vast design space of these alloys and (b) the need for numerous realizations to obtain a statistically accurate average due to the random solute environments. To address these limitations, machine learning (ML) and deep learning (DL) models offer a promising alternative by streamlining the computation of these input parameters, potentially reducing the reliance on computationally expensive brute-force methods.

The advent of ML and DL methods have revolutionized the ways of materials design, physical modeling, and properties measurement \cite{lecun2015deep, ramprasad2017machine, butler2018machine, wei2019machine, morgan2020opportunities, guo2021artificial}. These methods can discover hidden patterns in the training data and thus have been merged into all disciplines. The use of DL methods within the field of crystalline materials is wide, from machine learning interatomic potential development
\cite{behler2016perspective, shapeev2016moment, deringer2019machine, mueller2020machine, } to intrinsic properties calculation \cite{xie2018crystal} to dynamic crack path prediction \cite{hsu2020using}. Among all the DL architectures, GNN models have been developed to deal with graph structures which model a series of objects (nodes) and the relationships between them (edges), making them an ideal solution for crystalline materials where the atoms are nodes and the edges represent the intermediate bond \cite{yang2022linking, guo2022rapid}.
 
While machine learning (ML) methods have accelerated the exploration of multi-principal element alloys (MPEAs), they often target specific material properties in isolation. This narrow focus can limit their ability to incorporate the broader, interdisciplinary knowledge that is critical for true breakthroughs in alloy design and scientific discovery. To overcome these limitations, multi-agent systems have emerged as a transformative approach, facilitating the integration of multi-modal data and external knowledge—such as new developments in physics and materials science including theoretical models—into the design process in a more holistic and adaptive way \cite{ghafarollahi2024sciagentsautomatingscientificdiscovery, pei2024leveraging, guo2024large, xie2024large, cheng2024exploring, m2024augmenting, buehler2024generative, ni2024mechagents, stewart2024molecular, ghafarollahi2024protagents, ghafarollahi2024atomagents}. In multi-agent systems, a group of AI agents—each powered by Large Language Models (LLMs)—work together dynamically to solve complex, multi-faceted problems. By assigning distinct roles to each agent via targeted prompts, the system can divide and conquer different aspects of the alloy design challenge.

In this paper, we present an LLM-based multi-agent system that leverages the power of: (a) a suite of advanced Large Language Models (LLMs), responsible for fundamental tasks such as planning, reasoning, and decision-making; (b) a group of specialized agents, each playing distinct roles within the system; and (c) a set of external tools for various tasks, including a newly developed Graph Neural Network (GNN) model to predict atomic-level material properties. This multi-agent model builds upon AtomAgents, a multimodal multi-agent system with advanced simulation capabilities for alloy design and discovery. AtomAgents was capable of retrieving new physics from atomistic simulations, solving complex multi-faceted alloy discovery problems. However, as the complexity of the alloy systems increased, the computational cost of atomistic simulations became prohibitive.

To address this, we developed a GNN model that bypasses the limitations of brute-force calculations by providing rapid physics predictions. Although trained on a relatively small portion of the compositional space of the ternary Nb-Mo-Ta family of body-centered cubic (BCC) alloys, the GNN model accurately predicts fundamental quantities, including the Peierls barrier and solute/screw interaction energy. Several examples are provided to demonstrate the proficiency of this multi-agent approach in not only exploring the design space but also tackling complex alloy design challenges, such as predicting macroscopic yield stress.

The plan of the paper is as follows. We start by providing details about our GNN model to predict atomic-level material properties in Section \ref{sec:results}. We then discuss our proposed LLM-driven multi-agent system for automated mterials design, guided by the predictions from our GNN model. We outline the main components of the multi-agent approach and provide several examples to illustrate the proficiency of our model in solving complex alloy design problems. We then present the key findings in Section \ref{sec:conclusions} and discuss the implications of our multi-agent system for future research in materials discovery.

\section{Results and Discussion}\label{sec:results}
The main tool implemented in our multi-agent system is the GNN model that accurately predicts the fundamental material properties in multi-component alloys, i.e. Peierls barrier and potential energy change due to the screw dislocation movement. We first delve into the basics of development of the GNN model and then discuss our proposed LLM-based multi-agent system to design alloys solicitating physics data from the GNN model.
\subsection{GNN model}
\begin{figure}[ht!]
\centering
    \includegraphics[width=1\textwidth]{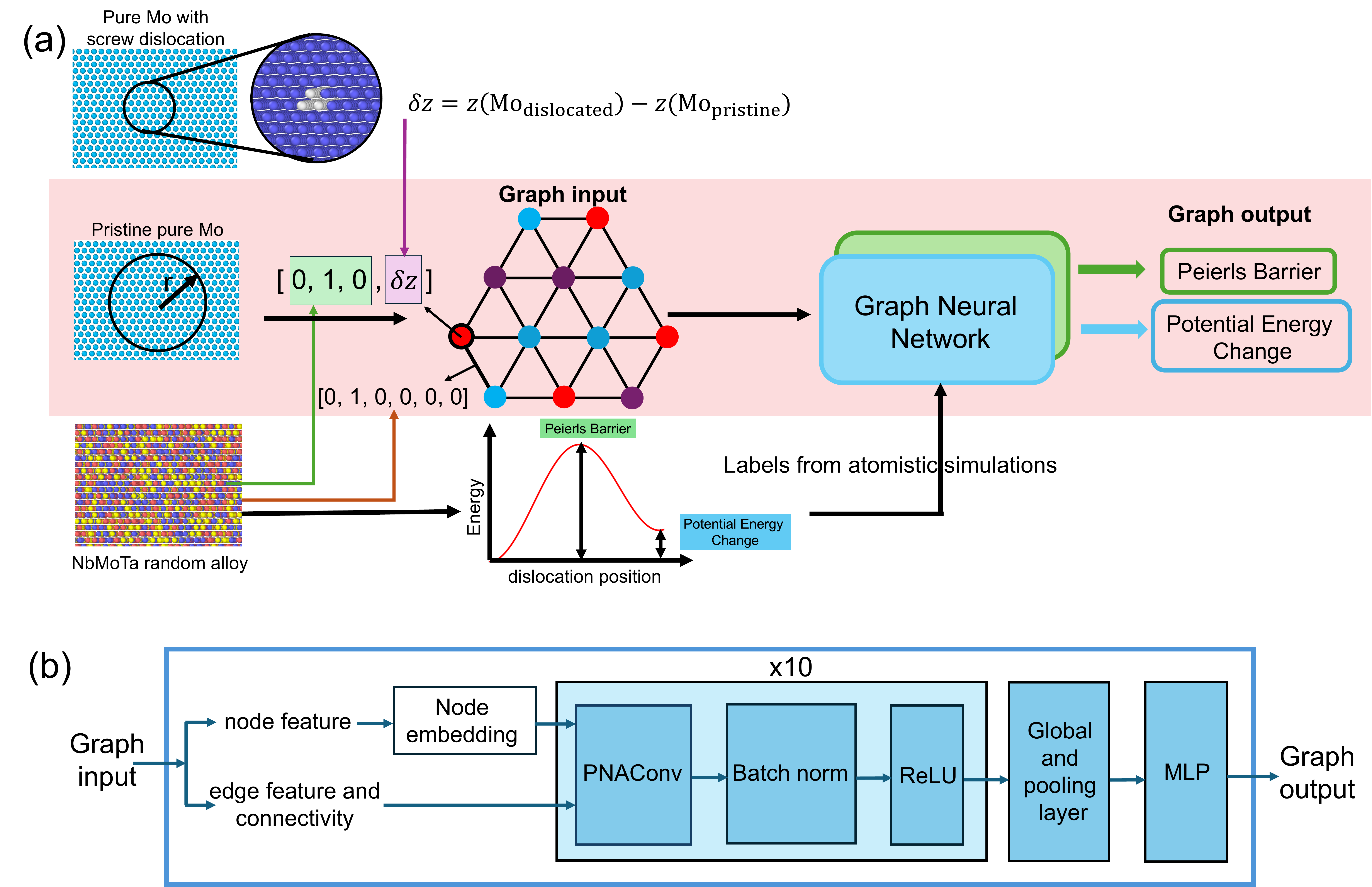}
    \caption{\textbf{(a) Overview of the workflow used in this work to train GNN models for an end-to-end prediction of the Peierls barrier and potential energy changes.} The process begins with the generation of initial and final dislocation structures in random BCC multi-component alloys, which are then minimized to compute the initial and final potential energies. Next, NEB simulations are performed to determine the minimum energy path and calculate the Peierls barrier. The input to the machine learning model is the graph representation of the alloys, with nodes encoding spatial and chemical information, and edges representing bond types. Two supervised GNN models are trained to predict graph-level labels: the Peierls barrier and potential energy change. (b) \textbf{GNN architecture.} The graph input is first passed into an input
block to enlarge the dimension of node features. The node embedding, edge features and connectivity of input graphs are then input to the message passing block where the information from the neighbors of each node in the graph is aggregated to update the hidden features of the nodes. The output of the message passing block is then input into a global pooling later which outputs a graph-level embedding by adding node embeddings across the node,  and it is connected to a multilayer perceptron (MLP) that returns a predicted graph label, i.e. Peierls barrier or potential energy change.}
    \label{fig:GNN_overview}
\end{figure}

The workflow of the GNN approach for predicting the Peierls barrier and energy changes due to dislocation movement is schematically presented in Figure \ref{fig:GNN_overview}. Initially, the dislocated structures for 200 samples per composition within the training compositional space are generated, as shown in Figure \ref{fig:ternary_composition} in the SI. After minimizing these structures, the potential energy change is calculated by subtracting the final potential energy from the initial one. Additionally, the Peierls barrier for each sample is extracted from the minimum energy path connecting the initial and final states, as derived from NEB simulations. The potential energy changes and the Peierls barriers serve as the ground truth or labels used to train the deep learning (DL) model. In this model, each random configuration is represented as a graph, where nodes correspond to atoms and edges to chemical bonds. Node features consist of chemical (solute types) and configurational features (screw dislocation displacement). The solute types at each site are one-hot encoded, with vectors $[1, 0, 0]$, $[0, 1, 0]$, and $[0, 0, 1]$ representing \textit{Nb}, \textit{Mo}, and \textit{Ta}, respectively. Additionally, the $z$-component of the screw dislocation displacement, $\delta_z$, is calculated for pure Mo and included as a node feature. Notably, only the chemical features vary across compositions, while the $\delta_z$ feature remains constant, thus eliminating the need for atomic relaxation in new random configurations during inference. Furthermore, edge features are constructed based on the solute types of the neighboring atoms. After constructing the graph input, the GNN model is trained to predict either the potential energy change or the Peierls barrier, which serves as the label for the entire graph. Once trained, the model can predict the potential energy change and Peierls barrier for new structures, drastically reducing the need for time-consuming atomistic simulations, particularly NEB calculations. This enables us to explore the full compositional space of NbMoTa ternary alloys to identify candidates with enhanced properties, such as a higher Peierls barrier. 

Peierls barrier and potential energy change predictions are graph regression tasks. In this work, we adopt a GNN with the Principal Neighborhood Aggregation graph convolution operator (PNAConv),\cite{corso2020principal} which has outperformed many popular GNN models in the literature, such as GCN,\cite{kipf2016semi} GAT,\cite{velickovic2017graph}, and GIN \cite{xu2018powerful} on benchmark graph regression tasks. The PNA model combines multiple aggregators, which affect how messages between nodes are passed, with a degree-scaler that generalizes the sum aggregator. Specifically, the architecture employs four different aggregators: \textit{mean}, \textit{maximum}, \textit{minimum}, and \textit{standard deviation}. The improved performance of the GNN is attributed to this combination, where the degree-based scalers amplify or attenuate signals in the network based on node degree.

More details about the construction of the dataset, the architecture of the GNN model, and the training procedure are provided in the Methods section. In the following sections, we evaluate the model's performance in predicting material properties on the test dataset and assess its ability to generalize to unseen configurations.

\subsubsection{Peierls barrier}
Random fluctuations in the solute environment surrounding the screw dislocation play a crucial role in determining the Peierls barrier, introducing a degree of complexity that must be captured for accurate predictions. In this section, we assess how well our GNN model predicts the Peierls barrier for various random compositions within the test set. To ensure the model's generalizability, the compositions in the test set are strategically distributed to cover a broad range of the compositional space, including the equimolar composition $\text{Nb}_{33}\text{Mo}_{33}\text{Ta}_{33}$ and binary systems. This comprehensive distribution allows us to verify that the model can accurately capture both the complexity of ternary alloys and the simpler binary ones, which is critical for extending the model to new, unseen compositions. Moreover, each composition consists of 50 realizations of random solute configurations, ensuring a robust evaluation of the model’s predictive capabilities across different atomic arrangements.

\begin{figure}[ht!]
\centering
    \includegraphics[width=1\textwidth]{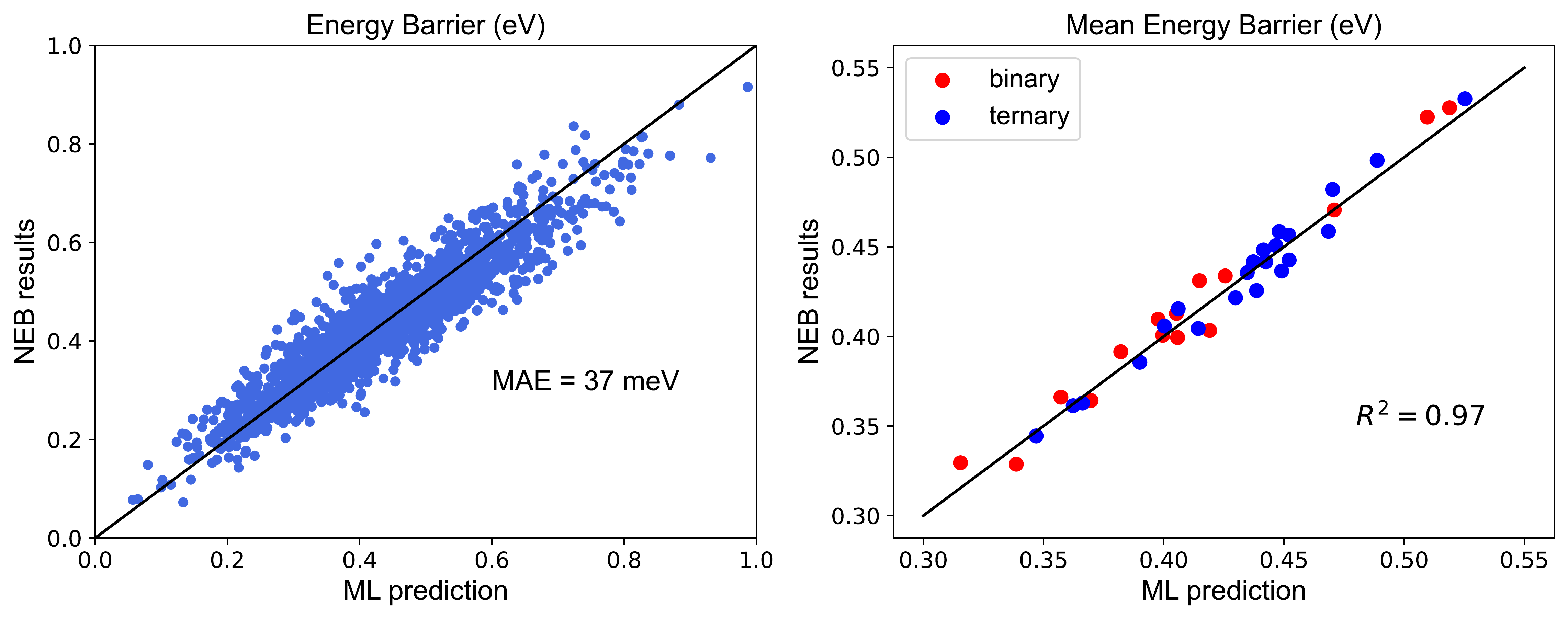}
    \caption{\textbf{Evaluation of the GNN model for Peierls barrier prediction on the test set.} Test set contains new compositions that never appeared in the training set. (a) Comparison of machine learning (ML) predictions and NEB results for the Peierls energy barrier. (b) Comparison of the ML results and NEB results for the mean energy barriers of binary and ternary compositions in the test set. The correlation coefficient $R^2$ between the predictions and the ground truth is 0.97.}
    \label{fig:barrier_test}
\end{figure}

We evaluate the model’s performance by comparing the predicted Peierls barrier values with the ground truth data, as illustrated in Figure \ref{fig:barrier_test}(a). The model achieves a low mean absolute error of 37 meV, indicating that it has effectively captured the fundamental physical mechanisms driving dislocation motion. This accuracy, despite the inherent randomness in the solute distribution, demonstrates the model’s ability to reliably predict the Peierls barrier across varying atomic environments. This result highlights the model's capacity to grasp complex atomic interactions that influence the Peierls barrier.

From these predictions, we can compute the mean energy barrier for each composition, a key parameter in solute strengthening theories, which rely on these barriers to predict material strength. Figure \ref{fig:barrier_test}(b) compares the mean Peierls barrier values predicted by the DL model against the ground truth values, showing a strong agreement across both ternary and binary alloys. 
This alignment highlights the model’s ability to predict individual configurations while also capturing broader trends in solute-strengthening behavior, offering valuable insights into how solute distributions influence material properties across various compositions.

In summary, the model’s low error and consistent performance across diverse alloy compositions affirm its potential as a powerful tool for predicting material behavior, significantly reducing the need for computationally expensive atomistic simulations such as NEB calculations.

\subsubsection{Potential energy change predictions}
\begin{figure}[ht!]
\centering
    \includegraphics[width=1\textwidth]{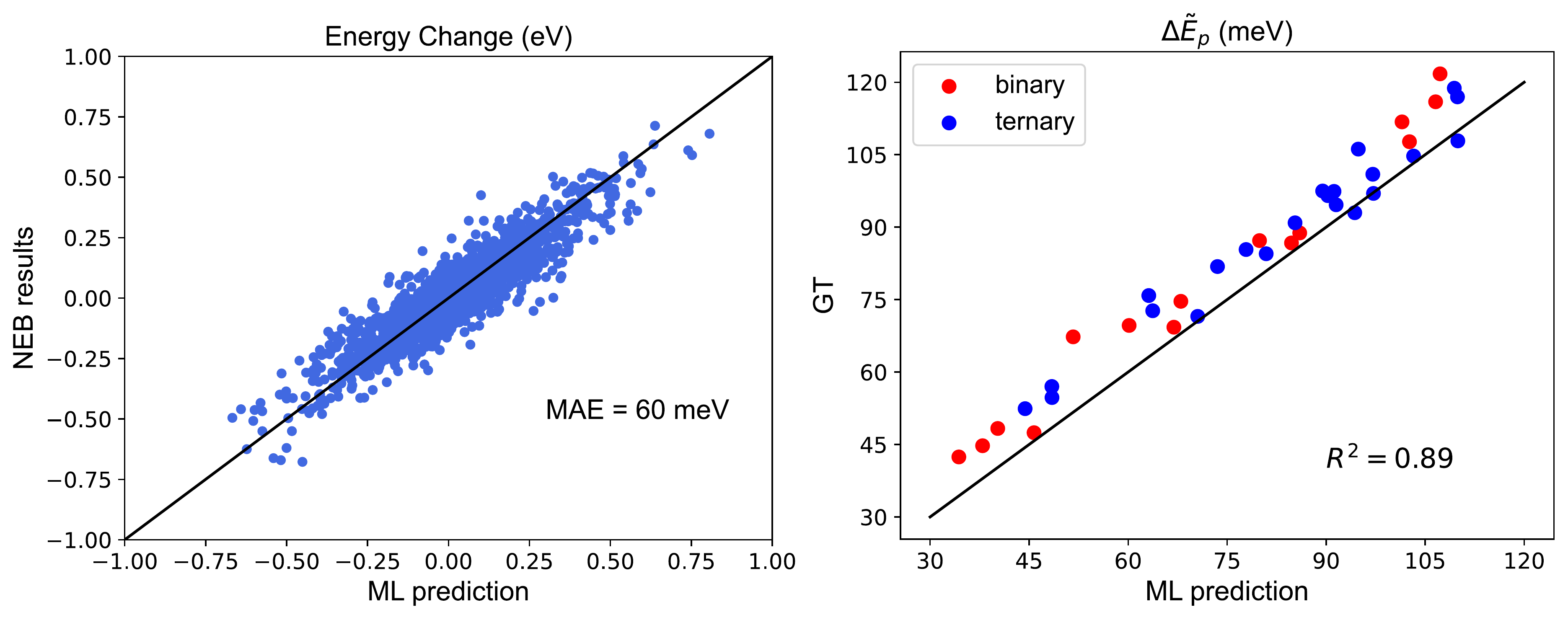}
    \caption{\textbf{Evaluation of the GNN model for the potential energy change prediction on the test set.} Test set contains new compositions that never appeared in the training set. (a) Comparison of ML predictions and NEB results for the potential energy change. (b) Comparison of the ML results and NEB results for the solute/screw interaction energy parameter, Eq. \ref{eg:dep} of binary and ternary compositions in the test set. The correlation coefficient $R^2$ between the predictions and the ground truth is 0.89.}
    \label{fig:energy_change_test}
\end{figure}

Another important quantity is the potential energy change as the dislocation moves from one energy minimum to the next. After training the GNN using the potential energy changes as graph labels, we compared the model’s predictions with the ground truth values for the graphs in the test set (see Fig. \ref{fig:ternary_composition}), as shown in Figure \ref{fig:energy_change_test}(a). The relatively low mean absolute error of 60 meV demonstrates the strong performance of our DL model in accurately predicting potential energy changes.

The random fluctuations in the solute environment lead to a distribution of energy changes. In random alloys, the average of this distribution is expected to be zero, and according to solute strengthening theories, the energy barrier against dislocation motion scales with the standard deviation of this distribution \cite{ghafarollahi2019solute, maresca2020theory}. This can be expressed as:

\begin{equation}\label{eg:dep}
\Delta \tilde{E}_{p}(a)=\left(\frac{b}{\zeta}\right)^{\frac{1}{2}}\sigma_{\Delta U}
\end{equation}
where $\zeta$ is the dislocation segment length (with $\zeta=5b$ in our case), and $b$ is the screw dislocation Burgers vector.

Figure \ref{fig:energy_change_test}(b) shows the values of the solute/screw interaction energy parameter computed by our DL model compared to the ground truth values, demonstrating good agreement for both ternary and binary alloys. This enables the rapid prediction of $\Delta \tilde{E}_{p}(a)$, a fundamental input parameter in solute-strengthening theories, providing critical insights for predicting material strength in multi-component alloys.

\subsection{Multi-modal multi-agent model}
\begin{figure}[t!]
\centering
    \includegraphics[width=.75\textwidth]{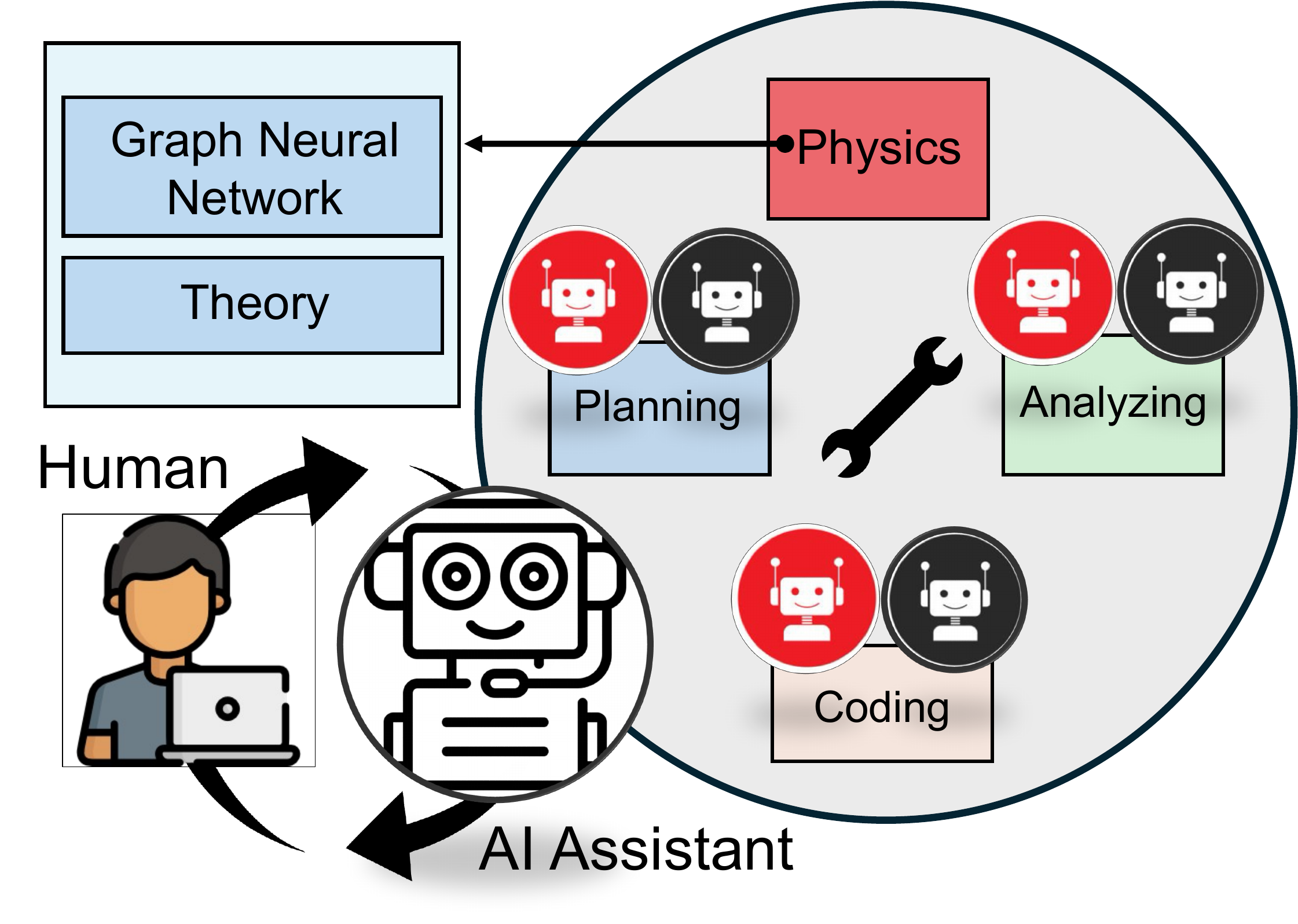}
    \caption{\textbf{Overview of the GNN-powered, LLM-based multi-agent system developed here.} The system consists of a human and an AI assistant agent at its core, where the human poses queries, and the AI assistant provides responses, seamlessly steering the problem-solving process with the help of integrated tools. These tools are responsible for various tasks, including planning, coding, and multi-modal analysis, and each incorporates a set of AI agents that dynamically collaborate to solve complex tasks. A key component is the physics tool, which includes newly developed graph neural network (GNN) models to retrieve essential physical parameters (such as the Peierls barrier and potential energy change) as well as physics-based theories (like solute-strengthening theory). The GNN models enable the rapid prediction of fundamental material properties, bypassing the need for costly atomistic simulations. The iterative collaboration between agents within the tools and the seamless interaction between the human and AI assistant allows for efficient resolution of complex materials design challenges.} 
    \label{fig:multiagent_model}
\end{figure}

Figure \ref{fig:multiagent_model} outlines our multi-agent model for rapid multi-component alloy design and analysis. Our model builds upon AtomAgents \cite{ghafarollahi2024atomagents}, a multi-modal multi-agent system designed to extract new physical insights from atomistic simulations. In this version, AtomAgents has been enhanced with our new graph neural network model that predicts fundamental physics data. This advancement offers an efficient way to explore the massive compositional space of multi-component alloys compared to costly atomistic simulations, allowing for solving more challenging problems and rapid design of alloys with enhanced or desired material properties. 

In the current multi-agent framework, the core agents are the "User" and the "AI Assistant," with the User posing the queries and the AI Assistant responding by utilizing various tools. These tools are designed to handle a wide range of tasks, from planning and coding to multi-modal data analysis. Additionally, these tools engage a set of autonomous agents that collaborate to respond effectively to the queries provided by the Assistant. A crucial component of this toolset is the physics tools, which empower the model by leveraging both newly developed Graph Neural Network (GNN) models and physics-based theoretical frameworks, such as solute-strengthening theories. These theories are essential for connecting atomistic-level features to macroscopic properties like yield stress. Each agent in our multi-agent system serves a specialized role, defined by a unique profile, and is powered by advanced general-purpose language models from the GPT family, accessed via the OpenAI API. A detailed description of each agent’s profile and the implementation details of the AI agents and tools are provided in Materials and Methods.

\begin{figure}[t!]
\centering
    \includegraphics[width=1\textwidth]{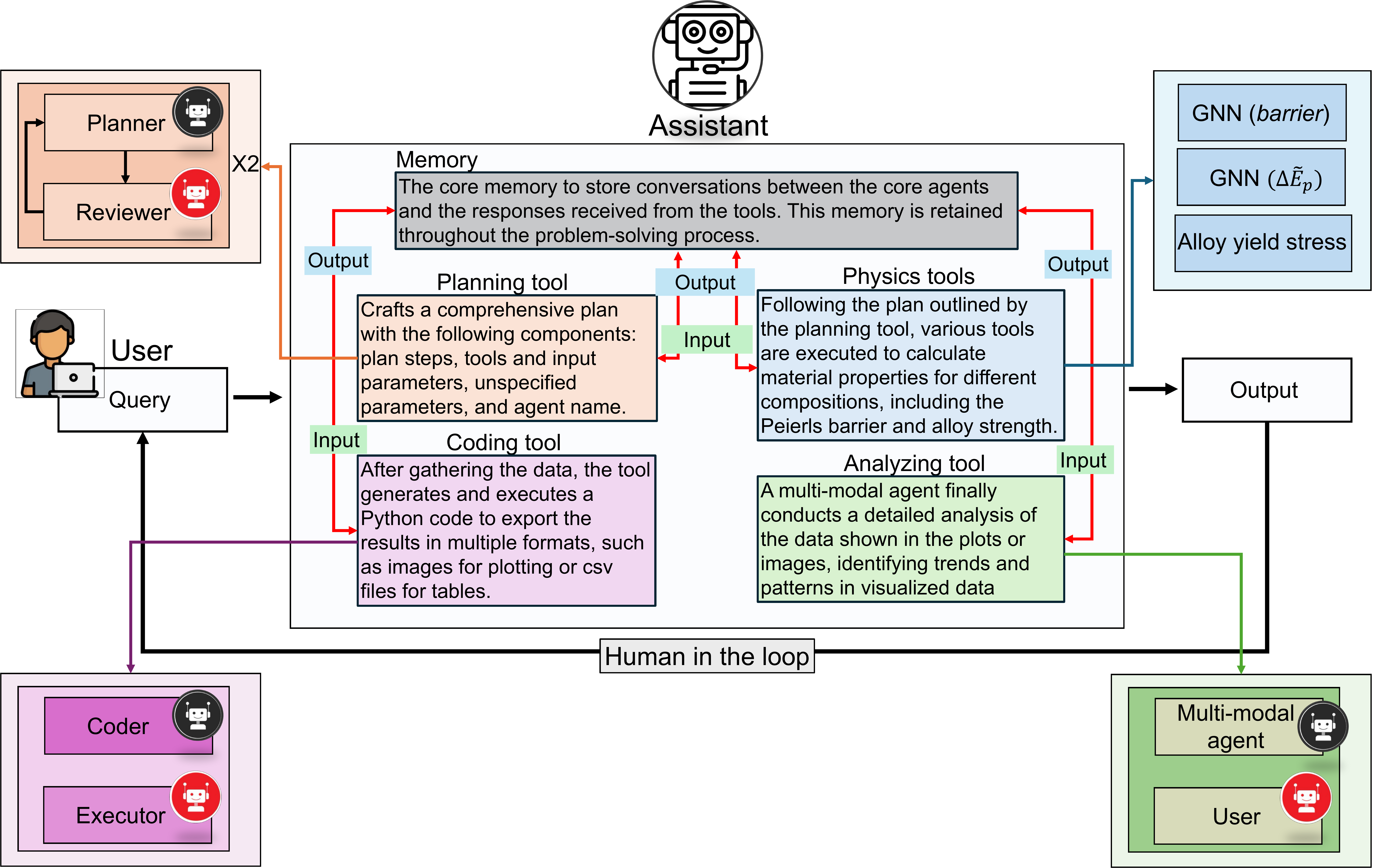}
    \caption{\textbf{Overview of a typical workflow executed by our multi-agent system for accelerated and automated multi-component alloy design and analysis.} Upon receiving a query from the user, the process begins with the assistant agent, which calls the planning tool to generate a detailed plan. The assistant agent follows this plan by invoking the relevant tools and providing the necessary input functions. These tools are integrated into our system to equip the agent with advanced capabilities, including material property predictions through physics-based retrieval tools (GNN models), data visualization via code execution (through a code-executor pair of agents), and multi-modal analysis (through a multi-modal agent-user pair) for enhanced visualization and reasoning. The system incorporates a human-in-the-loop design, fostering dynamic collaboration between humans and AI. This enables continuous refinement of ideas and queries, resulting in a more adaptive and efficient problem-solving process.}
    \label{fig:workflow_multiagent}
\end{figure}

The tools integrated within our multi-agent system enable the resolution of complex challenges in multi-component alloy design, allowing for automated exploration of the vast compositional design space to identify candidates with superior or targeted performance. These tools are automatically activated by the AI Assistant to perform various tasks throughout the alloy design and analysis process. A general workflow of our model is demonstrated in Figure \ref{fig:workflow_multiagent}. When the User submits a query, the Assistant calls the planning tool to generate a comprehensive plan, detailing the necessary steps, identifying the required tools, and specifying their input parameters. This planning process is driven by two key agents: the planner and the reviewer. The planner drafts the initial version of the plan, while the reviewer critiques and refines it through continuous iterations. This collaborative process ensures that the final plan is precise and thoroughly detailed, which is crucial as the entire problem-solving process depends on this plan.

Once the plan is finalized, the AI Assistant executes it, typically involving tasks such as predicting material properties using the physics tools, collecting and organizing data, and passing them to the coding tool for visualization and plotting. Finally, the Assistant engages the multi-modal agent to conduct a comprehensive analysis of the results.

The advantage of incorporating our newly developed GNN model into the multi-agent system is twofold: (a) it enables the rapid exploration of vast design spaces, providing deeper insights into material behavior across a wide range of compositions, and (b) it significantly accelerates the materials design process by allowing the fast computation of critical input parameters for physics-based theoretical models, such as solute-strengthening theories. To demonstrate the effectiveness of our multi-agent system in these key areas, we conducted two experiments, each detailed below.

\begin{figure}[t!]
\centering
    \includegraphics[width=0.85\textwidth]{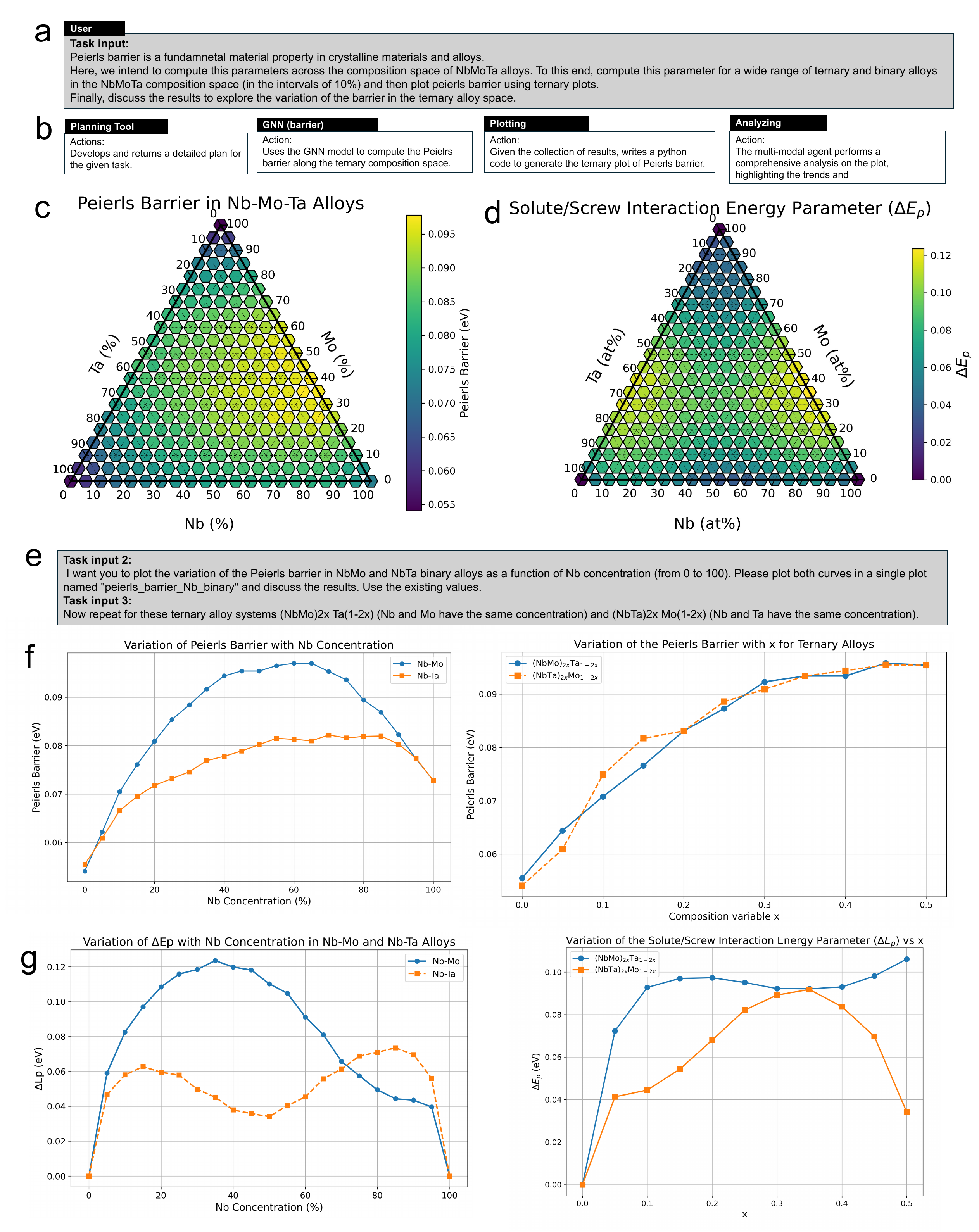}
    \caption{\textbf{Overview of the multi-agent collaboration to explore the variation of the Peierls barrier and solute/screw interaction energy parameter in binary and ternary alloys.} (a) User query requesting exploration of the Peierls barrier across a wide range of compositions, with a similar query repeated for the solute/screw interaction energy parameter. (b) Workflow detailing the computations and analyses performed by the multi-agent system. (c) and (d) Ternary plots displaying the Peierls barrier and the solute/screw interaction energy parameter, respectively, across the compositional space generated by the multi-agent system. (e) Follow-up tasks requested by the user, involving additional data plotting and analysis. (f) and (g) Plots illustrating the variation of the Peierls barrier and $\Delta \tilde{E_p}$, respectively, with solute concentration for binary and ternary alloys, generated by the multi-agent model in response to the follow-up tasks.}
    \label{fig:workflow_exp_barrier}
\end{figure}

\paragraph{Variation of the Peierls barrier in binary and ternary alloys}
In this experiment, the multi-agent system was tasked with exploring the Peierls barrier across the ternary compositional space, as illustrated in Figure \ref{fig:workflow_exp_barrier}(a). Upon receiving the query, the assistant agent activates the planning tool to devise a detailed plan for the task. The embedded planner and reviewer agents collaborate, with the planner drafting the initial plan and the reviewer providing feedback. This feedback loop iterates twice to ensure the accuracy and robustness of the plan. The resulting comprehensive plan outlines the following key steps: (a) Composition Generation: creating a list of all possible binary and ternary compositions in the Nb-Mo-Ta system at 5\% intervals (231 total compositions); (b) Computation of Peierls Barrier: using the GNN model to compute the Peierls barrier for all generated compositions; (c) Plotting Results: utilizing the coding tool to plot the computed Peierls barrier values on a ternary diagram representing the Nb-Mo-Ta compositional space; and (d) Data Analysis: applying the analysis tool to examine the plotted data for trends, correlations, and significant insights. The plan also includes a detailed list of functions and input parameters to ensure consistency and precision during execution.

The Assistant agent then executes the plan, generating the possible compositions in the ternary space and activating the relevant tools, as shown in Figure \ref{fig:workflow_exp_barrier}(b). We observe that the GNN model enables rapid predictions of the Peierls barrier across the entire compositional space (231 compositions at 5\% intervals). Notably, calculating the mean Peierls barrier for a single composition typically requires computationally expensive NEB simulations over numerous random configurations to capture the statistical nature of the problem. The GNN-based approach, however, completes these calculations within seconds per composition, demonstrating significant efficiency. After computing the Peierls barriers, the coding tool is called to write a Python function for plotting the values on a ternary diagram. Despite the complexity of ternary plots, the code generator, powered by the o1-mini LLM, produces the plot, as shown in Figure \ref{fig:workflow_exp_barrier}(c) , depicting the variation of the Peierls barrier across Nb-Mo-Ta ternary and binary compositions.

The results of the multi-agent system provide valuable insights into the variation of the Peierls barrier across the entire compositional space. Regions with high Peierls barrier values are identified around compositions with intermediate levels of Nb and Mo and low levels of Ta. As the final step of the workflow, the multi-modal agent performs detailed analysis, identifying the highest Peierls barrier in compositions with a high percentage of Mo and the lowest in those with a high percentage of Ta. Additionally, the agent conducts a comparative analysis of the influence of Nb, Mo, and Ta concentrations on the Peierls barrier.

However, some inconsistencies are observed in the agent’s analysis. For example, the model suggests that the Peierls barrier decreases with increasing Nb content, which is inaccurate based on the predictions. This highlights challenges in the multi-agent system’s ability to detect patterns in complex ternary plots, requiring further experimentation and refinement. The complete results of the ternary plot analysis are presented in Figure \ref{fig:plot_analysis_barrier_ternary} in the Supplementary Material.

The full spectrum of Peierls barrier values across the ternary compositional space offers many opportunities for further insights. For instance, it allows exploration of the barrier's variation within binary systems or specific ternary systems. The human-in-the-loop capability of the multi-agent system makes it possible to issue follow-up queries. Here, we instruct the model to plot the variation of the Peierls barrier with solute concentration for specific binary and ternary systems, including the Nb-Mo and Nb-Ta binary systems, as well as the (NbMo)$_{2x}$Ta$_{1-2x}$ and (NbTa)$_{2x}$Mo$_{1-2x}$ ternary systems, as shown in Figure \ref{fig:workflow_exp_barrier}(e).

The assistant agent identifies the relevant compositions from prior results and passes their Peierls barrier values to the coding agent for plotting, generating the plots shown in Figure \ref{fig:workflow_exp_barrier}(e). These plots reveal a non-linear variation of the Peierls barrier with solute concentration. Key insights include:
(a) In the Nb-Mo system, the Peierls barrier increases with Nb concentration, peaking around 50\%, before decreasing as Nb content approaches 100\%.
(b) In the Nb-Ta system, the Peierls barrier also increases with Nb concentration, but at a slower rate.
(c) In both ternary systems, the Peierls barrier shows an increasing trend as $x$ rises from 0 to 0.5.

The entire workflow is repeated for the solute/screw interaction energy parameter, $\Delta \tilde{E}{p}$, a key material property derived from potential energy changes. The problem-solving approach remains identical, but the model now uses the potential energy GNN to compute $\Delta \tilde{E}{p}$. The results are visualized in the ternary plot of Figure \ref{fig:workflow_exp_barrier}(d), showing non-linear variations of the interaction energy parameter with solute concentrations.

Further insights are gained by examining the variation of $\Delta \tilde{E}{p}$ for selected binary and ternary systems, as shown in Figure \ref{fig:workflow_exp_barrier}(g). The key findings include:
(a) In the Nb-Mo system, $\Delta \tilde{E}{p}$ increases with Nb concentration, peaking around 30\%, and then gradually decreases as Nb content continues to rise. This suggests that the interaction energy parameter is optimized at intermediate Nb concentrations.
(b) In the Nb-Ta system, $\Delta \tilde{E}{p}$ is higher at high Nb concentrations compared to the Nb-Mo system, indicating stronger interactions at these concentrations.
(c) In the (NbTa)$_{2x}$Mo$_{1-2x}$ system, $\Delta \tilde{E}{p}$ increases steadily with $x$ and decreases as Mo content becomes minimal.
(d) In the (NbMo)$_{2x}$Ta$_{1-2x}$ system, $\Delta \tilde{E}_{p}$ rises with increasing $x$, eventually stabilizing, suggesting a saturation point in interaction energy as Ta content diminishes.

These results highlight the complex, non-linear interactions between solute concentrations and the screw dislocation, which influence both the Peierls barrier and the solute/screw interaction energy parameter. The GNN-powered multi-agent system offers an effective framework for efficiently exploring the vast compositional space of multi-component alloys. It enables the identification of key trends and intricate behaviors, significantly enhancing the materials design process through faster and more intelligent exploration.

\paragraph{Yields stress in multi-component BCC alloys}
Solute-strengthening theories focus on the atomic-level mechanisms governing dislocation motion and their interactions with solutes, aiming to develop mechanistic models that predict temperature-dependent mechanical strength, specifically the yield stress, of materials. These models typically rely on a number of input parameters derived from first-principles calculations such as Density Functional Theory (DFT) or atomistic simulations based on empirical potentials. A prominent theory for screw dislocation strengthening in BCC alloys is the Maresca-Curtin theory \cite{maresca2020theory}, which requires inputs like the Peierls potential, the solute/screw interaction energy parameter, the kink formation energy, and vacancy/interstitial formation energies. However, due to the complex energy landscapes in multi-component systems, accurately computing these parameters poses a significant challenge. The newly developed GNN model addresses this by enabling the rapid and accurate computation of two critical parameters: the Peierls potential and the solute-dislocation interaction energy. This capability allows us to evaluate these parameters in mere seconds for each composition in the NbMoTa alloy space, offering a major breakthrough in efficiently predicting yield stress in multi-component alloys as a function of temperature.

\begin{figure}[t!]
\centering
    \includegraphics[width=1\textwidth]{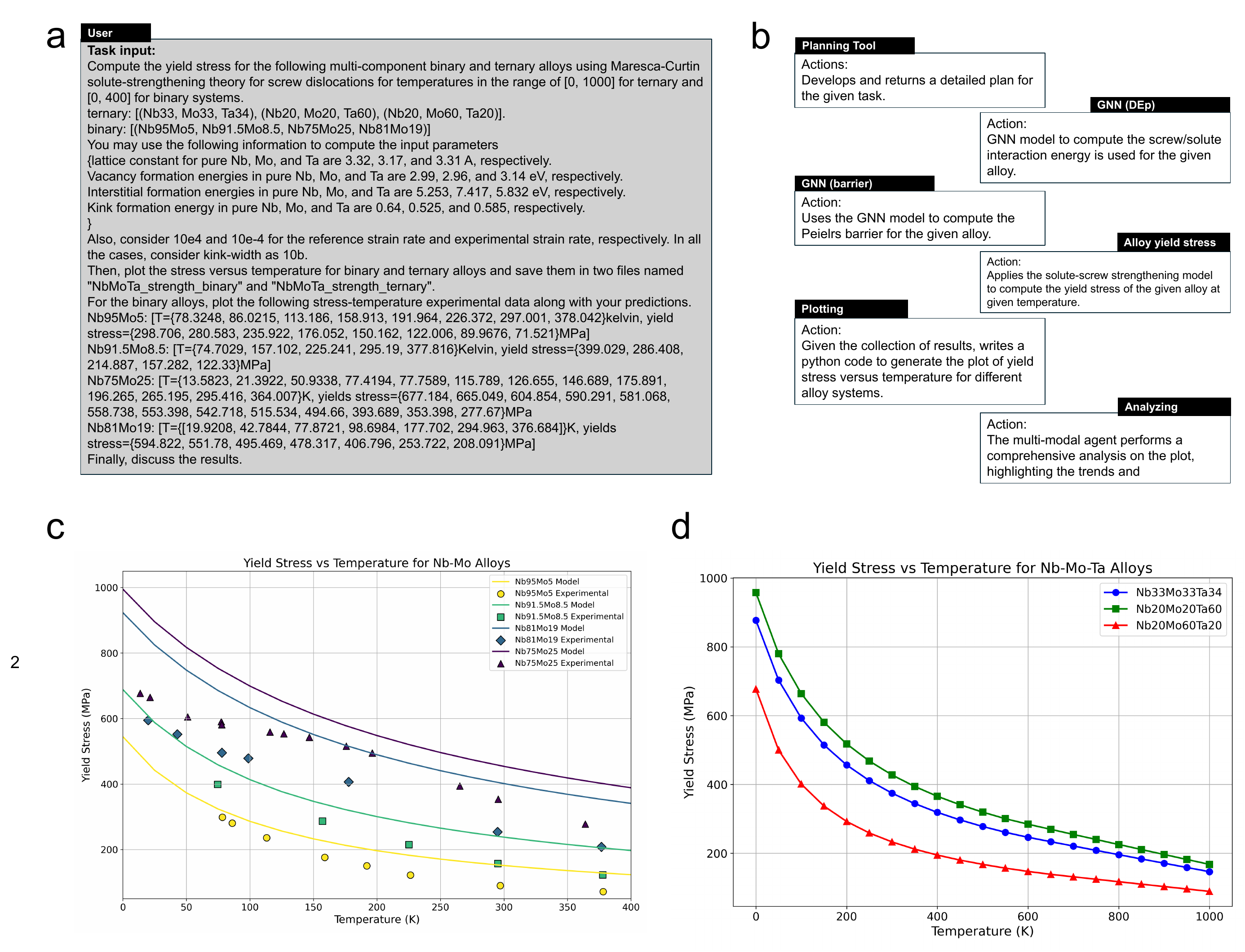}
    \caption{\textbf{Overview of the multi-agent collaboration to predict the yield stress in binary and ternary alloy BCC systems.} (a) The input task involving computing the yield stress for a set of compositions. Some material properties for the pure metals are provided as well as the strain rates. Moreover, the experimental data for binary alloys are provided to compare with the predictions. (b) The workflow of computations performed by our multi-agent system featuring key tasks like planning, tool calling for material property predictions from the GNN model, and finally plotting and analyzing the results. (c) Predictions of our model for the ternary alloys and (d) the predictions of our model for binary alloys along with the experimetal results.}
    \label{fig:workflow_exp_strength}
\end{figure}

The objective of this experiment is to demonstrate the potential of our multi-agent system in automating and significantly accelerating the materials design process by seamlessly integrating physics-based theoretical models with advanced DL-driven material prediction tools. In this task, the multi-agent system is assigned the calculation of yield stress for a series of alloy compositions over a broad range of temperatures, as shown in Figure \ref{fig:workflow_exp_strength}(a). Upon receiving the task from the User, the dynamic collaboration between agents is initiated to accomplish the goal, as illustrated in Figure \ref{fig:workflow_exp_strength}(b).

The process begins with the planning tool, which generates a detailed step-by-step plan outlining the required functions and their input parameters. The plan includes the use of GNN-powered tools to predict the Peierls barrier and the solute-dislocation interaction energy—key parameters for yield stress prediction. Additionally, for other input parameters such as the lattice constant, kink formation energy, and vacancy/interstitial formation energies, the system computes their values by averaging over the pure elements. Once all necessary input parameters are obtained, they are fed into the yield stress prediction tool for final calculations. The results are then passed to the coding tool, which generates Python code to plot the predictions. Finally, the multi-modal agent, powered by the GPT-4o LLM, analyzes the resulting plots to evaluate trends across compositions.

Despite the complexity of the task, which requires numerous calculations and input parameters, we observe that all sub-tasks are executed seamlessly. The primary outcomes of the multi-agent collaboration are presented in Figures \ref{fig:workflow_exp_strength}(c) and (d), which show the variation of yield stress with temperature for binary and ternary systems, respectively. The binary plot also includes experimental data from the input task for comparison.

Following this, the multi-modal agent conducts a comprehensive analysis of the results, as detailed in Figures \ref{fig:plot_analysis_strength_binary} and \ref{fig:plot_analysis_strength_ternary} in the Supplementary Information (SI), covering binary and ternary alloys, respectively. Upon completing the analysis, the assistant provides a summary of key insights, which is presented in Figure \ref{fig:plot_analysis_strength}.

The analysis indicates a typical decrease in alloy strength with increasing temperature, attributed to enhanced atomic mobility at higher temperatures. Additionally, the AI agent highlights that the Maresca-Curtin model predictions align with experimental data for some alloys, such as Nb81Mo19, but diverge for others, like Nb95Mo5 and Nb75Mo25. The agent suggests that these discrepancies may result from variations in alloy composition, microstructure, temperature effects, or experimental variability. To address these issues, the agent proposes refining the model to incorporate more detailed microstructural and compositional factors for improved accuracy.
Furthermore, the AI agent emphasizes the role of solid solution strengthening, with lattice distortions impeding dislocation motion as a key strengthening mechanism.

This multi-agent collaboration exemplifies the rigorous integration of multiple modalities, combining AI-driven physics predictions with established theoretical frameworks, along with plotting and visualization reasoning. This multifaceted approach not only enhances the precision and depth of the materials design process but also represents a transformative pathway that could revolutionize how materials design is approached and executed.

\begin{figure}[ht!]
    \centering
\begin{Box1}[colbacktitle={white!90!white}, colback={black!3!white}]{Plot analysis (GPT-4o)}
\footnotesize{\texttt{The analysis of the yield stress versus temperature plots for both binary and ternary Nb-Mo-Ta alloys provides valuable insights into the mechanical behavior of these materials.
\\
\#\#\# Binary Alloys:\\
- **General Trend**: The yield stress decreases with increasing temperature for all binary alloys, which is consistent with typical metallic behavior.\\
- **Model Accuracy**: The Maresca-Curtin model predictions align well with experimental data for some alloys, such as Nb81Mo19, but show discrepancies for others, like Nb95Mo5 and Nb75Mo25.\\
- **Discrepancies**: These may arise from variations in alloy composition, microstructural factors, temperature effects, and experimental variability.\\
- **Suggestions**: Refining the model to incorporate more detailed microstructural and compositional factors could improve accuracy.
\\
\#\#\# Ternary Alloys:\\
- **Yield Stress Variation**: The yield stress decreases with temperature for all ternary alloys. Nb20Mo20Ta60 exhibits the highest yield stress, while Nb20Mo60Ta20 shows the lowest.\\
- **Composition Effects**: High Ta content significantly enhances yield stress due to solid solution strengthening. Balanced compositions provide moderate strengthening, while high Mo content results in lower yield stress.\\
- **Strengthening Mechanisms**: Solid solution strengthening is the primary mechanism, with lattice distortions impeding dislocation motion. The effectiveness decreases with temperature due to increased atomic mobility.
\\
\#\#\# Overall Insights:\\
- **Design Implications**: Understanding the balance of alloying elements can aid in designing alloys with desired mechanical properties for specific applications.\\
- **Temperature Suitability**: These alloys may be more suitable for applications at lower temperatures where higher strength is required.
\\
The results highlight the importance of alloy composition and temperature in determining the mechanical properties of Nb-Mo-Ta alloys, providing a foundation for further research and development in this area.}}
\end{Box1}
    \caption{A summary of the analysis performed by the multi-modal agent of the yield stress versus temperature predictions as shown in Figures \ref{fig:workflow_exp_strength}c and d.}
    \label{fig:plot_analysis_strength}
\end{figure}

\section{Summary and future perspective}\label{sec:conclusions}
First, we developed a GNN model that directly links dislocated crystalline structures, caused by screw dislocations, to material properties in multi-component BCC alloy systems. Despite being trained on a small subset of the massive compositional space in ternary alloys, the model achieves high accuracy in predicting key properties. These include the potential energy change between consecutive screw dislocations and the Peierls barrier, which represents the highest energy at the transition state. One of the main applications of this energy change is the calculation of the solute/screw interaction energy parameter, a fundamental quantity in solute strengthening theories, which our model captures with high precision. The model’s predictions are generated in a matter of seconds, representing a significant improvement over conventional atomistic simulations, which can take days or even months when applied to the large design space of multi-component alloys. This speed, combined with minimal loss of accuracy, establishes our approach as a feasible alternative to costly simulations for investigating the effects of chemical fluctuations around defects on crystalline material properties. Furthermore, this method can be easily extended to high-entropy alloys and other crystalline structures, such as FCC and HCP systems, as well as to different types of defects, including edge dislocations. Our GNN model enables rapid exploration of the design space, paving the way for the development of alloys with enhanced mechanical properties, particularly those with higher strength.

Next, we constructed a multi-modal artificial intelligence (AI) system that integrates three core components: (a) Large Language Models (LLMs), which excel at a wide range of tasks, including multi-modal reasoning, strategic planning, rational thinking, and even coding; (b) AI agents, each powered by LLMs and external tools, designed with specialized roles and expertise, working collaboratively in a dynamic environment and autonomously tackling multi-faceted problems; and (c) a Graph Neural Network (GNN) model that provides rapid predictions of fundamental material properties. This multi-agent system is an enhanced version of our previous model, AtomAgents \cite{ghafarollahi2024atomagents}, which relied on direct atomistic simulations for physics-based insights. In the modified version, the GNN model is integrated to mitigate the high computational cost associated with atomistic simulations, thereby enabling faster exploration of the vast compositional design space inherent to multi-component alloys. This new framework offers a comprehensive method for addressing more complex problems in the design and analysis of multi-component alloys. Its capabilities are demonstrated through a series of experiments, including the exploration of the Peierls barrier and solute/screw interaction energy across the entire ternary composition space, as well as predictions of yield stress in BCC alloys, which are then validated against experimental data.

Our automated multi-agent system incorporates several key features that make it highly effective for complex materials science challenges. First, it is powered by advanced large language models (LLMs) like o1-preview and o1-mini, which provide exceptional capabilities in reasoning, strategic planning, and coding. These models excel at handling multi-modal data, evaluating emerging trends, and integrating information from diverse sources, making them essential for solving multi-scale problems. Second, the system supports the seamless integration of various data modalities, incorporating theoretical models, sophisticated deep learning algorithms, and even generative tools into the design and analysis process. This enables our system to stay at the cutting-edge of scientific and technological advancements, provideing a more holistic perspective on materials design. Third, the system is designed to support continuous evaluation and refinement, allowing for iterative problem-solving where errors are identified and corrected through mutual feedback between AI agents, ensuring that the system remains robust and adaptive throughout the problem-solving process. Finally, the human-in-the-loop functionality embedded within the system further enhances its flexibility and usability. This feature allows users to interact with the system directly, providing feedback, refining model parameters, or introducing additional requests. This AI-human collaboration is particularly valuable for complex, multi-faceted problems, ensuring that the system remains adaptable and responsive to specific user needs. Together, these advancements position the multi-agant modeling of materials as a powerful and innovative tool for tackling some of the most challenging problems in materials science and engineering, traditionally reserved for human experts.

Furthermore, the multi-agent system’s high degree of adaptability makes it suitable for application across a broad range of domains and disciplines, extending its potential impact beyond materials science. One of the primary aspects of this adaptability lies in the underlying LLM, which orchestrates the system’s internal operations and interactions. The performance of the entire system is heavily influenced by the capabilities of this LLM, and as LLM technology continues to evolve, we anticipate even greater improvements in system performance. For example, the transition from earlier versions to o1-preview resulted in significant gains in efficiency and accuracy. This ongoing advancement in LLMs suggests that future iterations of the system could become even more capable, especially as LLMs are further customized to specific tasks and integrated into different agents. Additionally, the system’s flexibility extends to its ability to incorporate a wide range of deep learning and generative models, each developed for various disciplines. This capability allows the system to address domain-specific challenges while maintaining a unified problem-solving framework. Another important feature of this system is its capacity to integrate with other multi-agent systems, such as SciAgents, \cite{ghafarollahi2024sciagentsautomatingscientificdiscovery, buehler2024graphreasoning} which specialize in generating advanced research hypotheses. This creates a cohesive scientific discovery process, where one system is responsible for exploring the hidden space of ideas, generating novel hypotheses, while another system is equipped to test, validate, or refine these ideas using advanced tools. Together, this interconnected approach provides a comprehensive and rigorous pathway for scientific discovery, making the multi-agent system an adaptable and powerful framework for tackling interdisciplinary problems.

\section{Materials and Methods}\label{sec:methods}

\subsection*{Dataset generation}
Random compositions of ternary NbMoTa refractory alloys with screw dislcoation are generated to compute the potential energy and Peierls barrier as follows. First pristine crystall strcuctures are generated by creating a rectangular simulation cell using the corresponding lattice constant and then randomly assigning solutes based on the desired concentrations. The simulation cell is oriented with dislocation glide direction x||[1,1,$\bar{2}$], glide plane normal direction, y||[$\bar{1}$10], and dislocation line direction, z||[111] with periodic boundary conditions along x and z and free surface along y. Atomic positions are relaxed by using a combination of the FIRE algorithm [97] and relaxation of the cell dimensions until the convergence is achieved-the norm of the force vector fell below 10e-6 eV/$\AA$ and stresses $\sigma_{xx}$ , $\sigma_{xz}$ , and $\sigma_{zz}$ fell below 0.1 MPa. All the atomistic simulations are performed using LAMMPS \cite{plimpton1995fast} at zero temperature. A machine learning MTP potential is used to describe the interatomic interactions in the NbMoTa alloys \cite{yin2021atomistic, shapeev2016moment}. 

We then introduce a screw dislocation in the center of each relaxed simulation cell using the PAD method \cite{ghafarollahi2020theory} and minimize the energy along with the relaxation of the pressures. This serves as our initial dislocated configuration. To generate the final dislocated configurations, we use the same initial pristine random structure and insert the dislocation at the adjacent position (next Peierls valley) relative to the initial dislocation at the distance of \textit{a} along the glide direction where $a$ is the Peierls valley distance. The potential energy changes is computed by subtracting the total potential energies of final and initial configurations. We then perform nudged elastic band (NEB) computations \cite{maras2016global, nakano2008space, henkelman2000climbing, henkelman2000improved} as implemented in LAMMPS \cite{plimpton1995fast} on these initial and final structures to compute the minimum energy path among these two energy states. The maximum value along this curve is stored as the corresponding Peierls barrier. The entire process is repeated for all the compositions in the training set (200 realizations for each composition) and test set (50 realizations for each composition) as depicted in Figure \ref{fig:ternary_composition}.  

\subsection*{Graph representation}
The results from atomistic simulations of atomic configurations and alloy properties are represented as graphs, which are used as inputs and labels to train the GNN. The graph structure is constructed from the pristine pure Mo configuration, where each Mo atom is a node in the graph, and the connectivity between nodes is determined by the distance between atoms. We consider a cylindrical region with a radius of $r_c=16\AA$, centered at the screw dislocation core, and only include atoms within this region as graph nodes. Within this graph, two atoms are connected if their distance is below the cutoff distance of 2.8 $\AA$, which is the distance between nearest neighbors in a perfect crystal of molybdenum. The graph representation consists of 705 nodes and 2610 edges.

The node features contain information about chemicals (solute type) and structural defect (screw dislocation). The structural defect feature is constructed from the $z$-component of the displacement derived from the atomistic simulation in pure Mo. This feature remains identical for all the atomic configurations in the dataset, eleminating the need for atomic relaxation for new configurations during inference. The chemical feature of the node is uniqely determined for each composition represented as the solute types converted to one-hot encoded representations ([1,0,0], [0,1,0], and [0,0,1] for Nb, Mo, Ta, respectively). Moreover, the edge features representing the bond type are encoded as one-hot representations based on the neighboring nodes as shown in Table \ref{tb: edge_features}

\begin{table*}[th!]
 \centering
\caption{Edge feature representation for bond type connecting neighboring nodes, $i$ and $j$.}
\begin{tabularx}{1.0\textwidth}{XXX}
\toprule 			
\textbf{node type i} & \textbf{node type j} & \textbf{edge feature}
\tabularnewline
\midrule
Nb & Nb & [1,0,0,0,0,0]
\\
Mo & Mo & [0,1,0,0,0,0]
\\
Ta & Ta & [0,0,1,0,0,0]
\\
Nb (Mo) & Mo (Ta) & [0,0,0,1,0,0]
\\
Nb (Ta) & Ta (Nb) & [0,0,0,0,1,0]
\\
Mo (Ta) & Ta (Mo) & [0,0,0,0,0,1]
\tabularnewline
\midrule
\end{tabularx}
\label{tb: edge_features}
\end{table*}

\subsection*{Graph neural networks (GNNs)}
The GNN model is developed based on the deep-learning framework PyTorch \cite{paszke2019pytorch} and its geometric extension library PyTorch Geometric \cite{fey2019fast}. The GNN architecture is depicted in Figure \ref{fig:GNN_overview}(b).  First of all, the input graph is sent to the input block which uses a combination of PNA convolutional (PNAConv) layer, Gated Recurrent Unit cell (GRUCell), and Batch Normalization layer (BatchNorm) to upscale the dimension of node features to 50. Then the graph is passed to the message passing block which contains 10-time repetition of the combined layers. Within the block, nodes communicate with each other by passing the messages given the node features plus connectivity and update their own node features considering the received messages. 
The last MLP that returns the predicted Peierls barrier or potential energy change has a structure of an input layer of size 30, a hidden layer of size 20, another hidden layer of size 10, and an output layer of one neuron. ReLU is adopted as the activation function in this MLP. We adopt default weight and bias initialization of all of the layers in the model defined by PyTorch Geometric.

\subsection*{Model training and evaluation}
All the datasets created for the training/validation compositions (see Figure  \ref{fig:ternary_composition}) are split into the train set (90\% data) and validation set (10\% data). The models were trained with a batch size of 32 using the Adam optimization method \cite{kingma2015adam} for 250 epochs on one NVIDIA Tesla V100s with 32GB memory. Training starts with a learning rate of 0.0005, and a dynamic learning rate scheduler named ReduceLROnPlateau reduces the learning rate by half if no improvement is seen for 10 epochs to minimize the validation MSE. The learning curves of the GNNs are shown in Fig. \ref{fig:loss_training_val} in the SI which indicate the convergence of training for both models.

\subsection*{Solute-strengthening theory in BCC alloys}
We apply Maresca-Curtin screw strengthening theory for non-dilute to high-entropy alloys to compute the yields stress \cite{maresca2020theory}. The theory is established based on the assumption that the initially straight dislocation becomes spontaneously kinked at zero load and zero temperature so as to lower their total energy. Three mechanisms contribute to the screw dislocation strengthening; (I) Peierls-like mechanism, (II) kink glide mechanism, and (III) cross-kink formation and unpinning and the alloy strength at temperature $T$ is expressed by
\begin{equation}
\tau(\dot{\epsilon}, T) = \tau_{xk}(\dot{\epsilon}, T) + \text{min}[\tau_k(\dot{\epsilon}, T), \tau_p(\dot{\epsilon}, T)]    
\end{equation}
where $\dot{\epsilon}$ is the experimental strain rate and $\tau_p$, $\tau_k$, and $\tau_{xk}$ are the Peierls strength, kink migration strength, and cross-kink unpinning strength, respectively. \cite{maresca2020theory, ghafarollahi2022screw}

\subsection*{Agent design}
We design AI agents using the state-of-the-art all-purpose LLM GPT-4 and dynamic multi-agent collaboration is implemented in AutoGen framework\cite{Zhu2023}, an open-source ecosystem for agent-based AI modeling. Additional agents are introduced as described below. 

In our multi-agent system, the human \textit{user} agent is constructed using UserProxyAgent class from Autogen, and \textit{Assistant}, \textit{Planner}, \textit{Reviewer}, \textit{coder} agents are created via AssistantAgent class from Autogen, while \textit{multi-modal} agent is constructed via MultimodalConversableAgent class. Each agent is assigned a role through a profile description 
as shown in Figures \ref{fig:profile_assistant}-\ref{fig:multi_modal_agent} in the Supplementary Material, included as \textit{system\_message} at their creation.   

\subsection*{Function and tool design}
All the tools implemented in this work are defined as python functions. Each function is characterized by a name, a description, and input properties. The full list of tools and their descriptions can be found in the corresponding codes.

\subsection*{Data and code availability}
All data and codes are available on GitHub at 
\url{https://github.com/lamm-mit/AlloyAgents}
. Alternatively, they will be provided by the corresponding author based on reasonable request.  
\\
\\
\textbf{Author Contributions:}
M.J.B and A.G. conceived the overall concept. A.G and M.J.B developed the GNN model and multi-agent system. A.G. curated the training and testing data for the GNN model, performed the tests for various problems, analyzed the results and prepared the first draft of the paper. M.J.B supported the analysis, revised and finalized the paper with A.G. 

\subsection*{Supplementary Materials}
Additional materials are provided as Supplementary Materials. 

\section*{Acknowledgements}
We acknowledge support from USDA (2021-69012-35978), DOE-SERDP (WP22-S1-3475), ARO (79058LSCSB, W911NF-22-2-0213 and W911NF2120130), and MIT’s Generative AI Initiative. A.G. gratefully acknowledges the financial support from the Swiss National Science Foundation (\#P500PT\_214448).


\bibliographystyle{unsrt}  
\bibliography{library} 

\begin{thebibliography}{10}

\bibitem{ye2016high}
YF~Ye, Qing Wang, Jiatian Lu, CT~Liu, and Yancong Yang.
\newblock High-entropy alloy: challenges and prospects.
\newblock {\em Materials Today}, 19(6):349--362, 2016.

\bibitem{miracle2017critical}
Daniel~B Miracle and Oleg~N Senkov.
\newblock A critical review of high entropy alloys and related concepts.
\newblock {\em Acta materialia}, 122:448--511, 2017.

\bibitem{george2019high}
Easo~P George, Dierk Raabe, and Robert~O Ritchie.
\newblock High-entropy alloys.
\newblock {\em Nature reviews materials}, 4(8):515--534, 2019.

\bibitem{george2020high}
Easo~P George, William~A Curtin, and Cemal~Cem Tasan.
\newblock High entropy alloys: A focused review of mechanical properties and deformation mechanisms.
\newblock {\em Acta Materialia}, 188:435--474, 2020.

\bibitem{senkov2010refractory}
ON~Senkov, GB~Wilks, DB~Miracle, CP~Chuang, and PK~Liaw.
\newblock Refractory high-entropy alloys.
\newblock {\em Intermetallics}, 18(9):1758--1765, 2010.

\bibitem{senkov2011mechanical}
Oleg~N Senkov, Garth~B Wilks, James~M Scott, and Daniel~B Miracle.
\newblock Mechanical properties of nb25mo25ta25w25 and v20nb20mo20ta20w20 refractory high entropy alloys.
\newblock {\em Intermetallics}, 19(5):698--706, 2011.

\bibitem{senkov2011microstructure}
ON~Senkov, JM~Scott, SV~Senkova, DB~Miracle, and CF~Woodward.
\newblock Microstructure and room temperature properties of a high-entropy tanbhfzrti alloy.
\newblock {\em Journal of alloys and compounds}, 509(20):6043--6048, 2011.

\bibitem{senkov2018development}
Oleg~N Senkov, Daniel~B Miracle, Kevin~J Chaput, and Jean-Philippe Couzinie.
\newblock Development and exploration of refractory high entropy alloys—a review.
\newblock {\em Journal of materials research}, 33(19):3092--3128, 2018.

\bibitem{senkov2018compositional}
ON~Senkov, S~Rao, KJ~Chaput, and C~Woodward.
\newblock Compositional effect on microstructure and properties of nbtizr-based complex concentrated alloys.
\newblock {\em Acta Materialia}, 151:201--215, 2018.

\bibitem{han2018microstructures}
ZD~Han, HW~Luan, X~Liu, N~Chen, XY~Li, Y~Shao, and KF~Yao.
\newblock Microstructures and mechanical properties of tixnbmotaw refractory high-entropy alloys.
\newblock {\em Materials Science and Engineering: A}, 712:380--385, 2018.

\bibitem{senkov2019high}
Oleg~N Senkov, St{\'e}phane Gorsse, and Daniel~B Miracle.
\newblock High temperature strength of refractory complex concentrated alloys.
\newblock {\em Acta materialia}, 175:394--405, 2019.

\bibitem{xiong2023refractory}
Wei Xiong, Amy~XY Guo, Shuai Zhan, Chain-Tsuan Liu, and Shan~Cecilia Cao.
\newblock Refractory high-entropy alloys: A focused review of preparation methods and properties.
\newblock {\em Journal of Materials Science \& Technology}, 142:196--215, 2023.

\bibitem{varvenne2016theory}
C{\'e}line Varvenne, Aitor Luque, and William~A Curtin.
\newblock Theory of strengthening in fcc high entropy alloys.
\newblock {\em Acta Materialia}, 118:164--176, 2016.

\bibitem{varvenne2017solute}
C{\'e}line Varvenne, Gerard Paul~M Leyson, Maryam Ghazisaeidi, and William~A Curtin.
\newblock Solute strengthening in random alloys.
\newblock {\em Acta Materialia}, 124:660--683, 2017.

\bibitem{rao2019solution}
SI~Rao, E~Antillon, C~Woodward, B~Akdim, TA~Parthasarathy, and ON~Senkov.
\newblock Solution hardening in body-centered cubic quaternary alloys interpreted using suzuki's kink-solute interaction model.
\newblock {\em Scripta Materialia}, 165:103--106, 2019.

\bibitem{ghafarollahi2020theory}
Alireza Ghafarollahi and William~A Curtin.
\newblock Theory of double-kink nucleation in dilute bcc alloys.
\newblock {\em Acta Materialia}, 196:635--650, 2020.

\bibitem{ghafarollahi2021theory}
Alireza Ghafarollahi and WA~Curtin.
\newblock Theory of kink migration in dilute bcc alloys.
\newblock {\em Acta Materialia}, 215:117078, 2021.

\bibitem{ghamaresca2020mechanistic}
Francesco Maresca and William~A. Curtin.
\newblock Mechanistic origin of high strength in refractory bcc high entropy alloys up to 1900k.
\newblock {\em Acta Materialia}, 182:235--249, 2020.

\bibitem{rao2021theory}
SI~Rao, C~Woodward, B~Akdim, Oleg~N Senkov, and D~Miracle.
\newblock Theory of solid solution strengthening of bcc chemically complex alloys.
\newblock {\em Acta Materialia}, 209:116758, 2021.

\bibitem{kubilay2021high}
RE~Kubilay, A~Ghafarollahi, F~Maresca, and WA~Curtin.
\newblock High energy barriers for edge dislocation motion in body-centered cubic high entropy alloys.
\newblock {\em Npj computational materials}, 7(1):112, 2021.

\bibitem{ghafarollahi2022screw}
Alireza Ghafarollahi and William~A Curtin.
\newblock Screw-controlled strength of bcc non-dilute and high-entropy alloys.
\newblock {\em Acta Materialia}, 226:117617, 2022.

\bibitem{baruffi2022screw}
C~Baruffi, F~Maresca, and WA~Curtin.
\newblock Screw vs. edge dislocation strengthening in body-centered-cubic high entropy alloys and implications for guided alloy design.
\newblock {\em Mrs Communications}, 12(6):1111--1118, 2022.

\bibitem{rao2022theory}
Y~Rao, C~Baruffi, A~De~Luca, C~Leinenbach, and WA~Curtin.
\newblock Theory-guided design of high-strength, high-melting point, ductile, low-density, single-phase bcc high entropy alloys.
\newblock {\em Acta Materialia}, 237:118132, 2022.

\bibitem{maresca2020theory}
Francesco Maresca and William~A Curtin.
\newblock Theory of screw dislocation strengthening in random bcc alloys from dilute to “high-entropy” alloys.
\newblock {\em Acta Materialia}, 182:144--162, 2020.

\bibitem{maresca2020mechanistic}
Francesco Maresca and William~A Curtin.
\newblock Mechanistic origin of high strength in refractory bcc high entropy alloys up to 1900k.
\newblock {\em Acta Materialia}, 182:235--249, 2020.

\bibitem{ghafarollahi2019solute}
A~Ghafarollahi, F~Maresca, and WA~Curtin.
\newblock Solute/screw dislocation interaction energy parameter for strengthening in bcc dilute to high entropy alloys.
\newblock {\em Modelling and Simulation in Materials Science and Engineering}, 27(8):085011, 2019.

\bibitem{lecun2015deep}
Yann LeCun, Yoshua Bengio, and Geoffrey Hinton.
\newblock Deep learning.
\newblock {\em nature}, 521(7553):436--444, 2015.

\bibitem{ramprasad2017machine}
Rampi Ramprasad, Rohit Batra, Ghanshyam Pilania, Arun Mannodi-Kanakkithodi, and Chiho Kim.
\newblock Machine learning in materials informatics: recent applications and prospects.
\newblock {\em npj Computational Materials}, 3(1):54, 2017.

\bibitem{butler2018machine}
Keith~T Butler, Daniel~W Davies, Hugh Cartwright, Olexandr Isayev, and Aron Walsh.
\newblock Machine learning for molecular and materials science.
\newblock {\em Nature}, 559(7715):547--555, 2018.

\bibitem{wei2019machine}
Jing Wei, Xuan Chu, Xiang-Yu Sun, Kun Xu, Hui-Xiong Deng, Jigen Chen, Zhongming Wei, and Ming Lei.
\newblock Machine learning in materials science.
\newblock {\em InfoMat}, 1(3):338--358, 2019.

\bibitem{morgan2020opportunities}
Dane Morgan and Ryan Jacobs.
\newblock Opportunities and challenges for machine learning in materials science.
\newblock {\em Annual Review of Materials Research}, 50(1):71--103, 2020.

\bibitem{guo2021artificial}
Kai Guo, Zhenze Yang, Chi-Hua Yu, and Markus~J Buehler.
\newblock Artificial intelligence and machine learning in design of mechanical materials.
\newblock {\em Materials Horizons}, 8(4):1153--1172, 2021.

\bibitem{behler2016perspective}
J{\"o}rg Behler.
\newblock Perspective: Machine learning potentials for atomistic simulations.
\newblock {\em The Journal of chemical physics}, 145(17), 2016.

\bibitem{shapeev2016moment}
Alexander~V Shapeev.
\newblock Moment tensor potentials: A class of systematically improvable interatomic potentials.
\newblock {\em Multiscale Modeling \& Simulation}, 14(3):1153--1173, 2016.

\bibitem{deringer2019machine}
Volker~L Deringer, Miguel~A Caro, and G{\'a}bor Cs{\'a}nyi.
\newblock Machine learning interatomic potentials as emerging tools for materials science.
\newblock {\em Advanced Materials}, 31(46):1902765, 2019.

\bibitem{mueller2020machine}
Tim Mueller, Alberto Hernandez, and Chuhong Wang.
\newblock Machine learning for interatomic potential models.
\newblock {\em The Journal of chemical physics}, 152(5), 2020.

\bibitem{}
Markus~J Buehler.
\newblock Melm, a generative pretrained language modeling framework that solves forward and inverse mechanics problems.

\bibitem{xie2018crystal}
Tian Xie and Jeffrey~C Grossman.
\newblock Crystal graph convolutional neural networks for an accurate and interpretable prediction of material properties.
\newblock {\em Physical review letters}, 120(14):145301, 2018.

\bibitem{hsu2020using}
Yu-Chuan Hsu, Chi-Hua Yu, and Markus~J Buehler.
\newblock Using deep learning to predict fracture patterns in crystalline solids.
\newblock {\em Matter}, 3(1):197--211, 2020.

\bibitem{yang2022linking}
Zhenze Yang and Markus~J Buehler.
\newblock Linking atomic structural defects to mesoscale properties in crystalline solids using graph neural networks.
\newblock {\em Npj Computational Materials}, 8(1):198, 2022.

\bibitem{guo2022rapid}
Kai Guo and Markus~J Buehler.
\newblock Rapid prediction of protein natural frequencies using graph neural networks.
\newblock {\em Digital Discovery}, 1(3):277--285, 2022.

\bibitem{ghafarollahi2024sciagentsautomatingscientificdiscovery}
Alireza Ghafarollahi and Markus~J. Buehler.
\newblock Sciagents: Automating scientific discovery through multi-agent intelligent graph reasoning.
\newblock 2024.

\bibitem{pei2024leveraging}
Qizhi Pei, Lijun Wu, Kaiyuan Gao, Jinhua Zhu, Yue Wang, Zun Wang, Tao Qin, and Rui Yan.
\newblock Leveraging biomolecule and natural language through multi-modal learning: A survey.
\newblock {\em arXiv preprint arXiv:2403.01528}, 2024.

\bibitem{guo2024large}
Taicheng Guo, Xiuying Chen, Yaqi Wang, Ruidi Chang, Shichao Pei, Nitesh~V Chawla, Olaf Wiest, and Xiangliang Zhang.
\newblock Large language model based multi-agents: A survey of progress and challenges.
\newblock {\em arXiv preprint arXiv:2402.01680}, 2024.

\bibitem{xie2024large}
Junlin Xie, Zhihong Chen, Ruifei Zhang, Xiang Wan, and Guanbin Li.
\newblock Large multimodal agents: A survey.
\newblock {\em arXiv preprint arXiv:2402.15116}, 2024.

\bibitem{cheng2024exploring}
Yuheng Cheng, Ceyao Zhang, Zhengwen Zhang, Xiangrui Meng, Sirui Hong, Wenhao Li, Zihao Wang, Zekai Wang, Feng Yin, Junhua Zhao, et~al.
\newblock Exploring large language model based intelligent agents: Definitions, methods, and prospects.
\newblock {\em arXiv preprint arXiv:2401.03428}, 2024.

\bibitem{m2024augmenting}
Andres M.~Bran, Sam Cox, Oliver Schilter, Carlo Baldassari, Andrew~D White, and Philippe Schwaller.
\newblock Augmenting large language models with chemistry tools.
\newblock {\em Nature Machine Intelligence}, pages 1--11, 2024.

\bibitem{buehler2024generative}
Markus~J Buehler.
\newblock Generative retrieval-augmented ontologic graph and multiagent strategies for interpretive large language model-based materials design.
\newblock {\em ACS Engineering Au}, 4(2):241--277, 2024.

\bibitem{ni2024mechagents}
Bo~Ni and Markus~J Buehler.
\newblock Mechagents: Large language model multi-agent collaborations can solve mechanics problems, generate new data, and integrate knowledge.
\newblock {\em Extreme Mechanics Letters}, 67:102131, 2024.

\bibitem{stewart2024molecular}
Isabella Stewart and Markus Buehler.
\newblock Molecular analysis and design using multimodal generative artificial intelligence via multi-agent modeling.
\newblock 2024.

\bibitem{ghafarollahi2024protagents}
Alireza Ghafarollahi and Markus~J Buehler.
\newblock Protagents: protein discovery via large language model multi-agent collaborations combining physics and machine learning.
\newblock {\em Digital Discovery}, 2024.

\bibitem{ghafarollahi2024atomagents}
Alireza Ghafarollahi and Markus~J Buehler.
\newblock Atomagents: Alloy design and discovery through physics-aware multi-modal multi-agent artificial intelligence.
\newblock {\em arXiv preprint arXiv:2407.10022}, 2024.

\bibitem{corso2020principal}
Gabriele Corso, Luca Cavalleri, Dominique Beaini, Pietro Li{\`o}, and Petar Veli{\v{c}}kovi{\'c}.
\newblock Principal neighbourhood aggregation for graph nets.
\newblock {\em Advances in Neural Information Processing Systems}, 33:13260--13271, 2020.

\bibitem{kipf2016semi}
Thomas~N Kipf and Max Welling.
\newblock Semi-supervised classification with graph convolutional networks.
\newblock {\em arXiv preprint arXiv:1609.02907}, 2016.

\bibitem{velickovic2017graph}
Petar Velickovic, Guillem Cucurull, Arantxa Casanova, Adriana Romero, Pietro Lio, Yoshua Bengio, et~al.
\newblock Graph attention networks.
\newblock {\em stat}, 1050(20):10--48550, 2017.

\bibitem{xu2018powerful}
Keyulu Xu, Weihua Hu, Jure Leskovec, and Stefanie Jegelka.
\newblock How powerful are graph neural networks?
\newblock {\em arXiv preprint arXiv:1810.00826}, 2018.

\bibitem{buehler2024graphreasoning}
Markus~J. Buehler.
\newblock Accelerating scientific discovery with generative knowledge extraction, graph-based representation, and multimodal intelligent graph reasoning.
\newblock {\em Machine Learning: Science and Technology}, 2024.

\bibitem{plimpton1995fast}
Steve Plimpton.
\newblock Fast parallel algorithms for short-range molecular dynamics.
\newblock {\em Journal of computational physics}, 117(1):1--19, 1995.

\bibitem{yin2021atomistic}
Sheng Yin, Yunxing Zuo, Anas Abu-Odeh, Hui Zheng, Xiang-Guo Li, Jun Ding, Shyue~Ping Ong, Mark Asta, and Robert~O Ritchie.
\newblock Atomistic simulations of dislocation mobility in refractory high-entropy alloys and the effect of chemical short-range order.
\newblock {\em Nature communications}, 12(1):4873, 2021.

\bibitem{maras2016global}
Emile Maras, Oleg Trushin, Alexander Stukowski, Tapio Ala-Nissila, and Hannes Jonsson.
\newblock Global transition path search for dislocation formation in ge on si (001).
\newblock {\em Computer Physics Communications}, 205:13--21, 2016.

\bibitem{nakano2008space}
Aiichiro Nakano.
\newblock A space--time-ensemble parallel nudged elastic band algorithm for molecular kinetics simulation.
\newblock {\em Computer Physics Communications}, 178(4):280--289, 2008.

\bibitem{henkelman2000climbing}
Graeme Henkelman, Blas~P Uberuaga, and Hannes J{\'o}nsson.
\newblock A climbing image nudged elastic band method for finding saddle points and minimum energy paths.
\newblock {\em The Journal of chemical physics}, 113(22):9901--9904, 2000.

\bibitem{henkelman2000improved}
Graeme Henkelman and Hannes J{\'o}nsson.
\newblock Improved tangent estimate in the nudged elastic band method for finding minimum energy paths and saddle points.
\newblock {\em The Journal of chemical physics}, 113(22):9978--9985, 2000.

\bibitem{paszke2019pytorch}
Adam Paszke, Sam Gross, Francisco Massa, Adam Lerer, James Bradbury, Gregory Chanan, Trevor Killeen, Zeming Lin, Natalia Gimelshein, Luca Antiga, et~al.
\newblock Pytorch: An imperative style, high-performance deep learning library.
\newblock {\em Advances in neural information processing systems}, 32, 2019.

\bibitem{fey2019fast}
Matthias Fey and Jan~Eric Lenssen.
\newblock Fast graph representation learning with pytorch geometric.
\newblock {\em arXiv preprint arXiv:1903.02428}, 2019.

\bibitem{kingma2015adam}
Diederik~P Kingma and JL~Ba.
\newblock Adam: A method for stochastic optimization 3rd international conference on learning representations.
\newblock In {\em ICLR 2015-Conference Track Proceedings}, volume~1, 2015.

\bibitem{Zhu2023}
Chenxu Zhu, Bo~Chen, Huifeng Guo, Hang Xu, Xiangyang Li, Xiangyu Zhao, Weinan Zhang, Yong Yu, and Ruiming Tang.
\newblock Autogen: An automated dynamic model generation framework for recommender system.
\newblock {\em WSDM 2023 - Proceedings of the 16th ACM International Conference on Web Search and Data Mining}, pages 598--606, 2 2023.

\end{thebibliography}

\newpage
\appendix

\pagestyle{empty} 

\renewcommand{\thefigure}{S\arabic{figure}}
\setcounter{figure}{0} 
\renewcommand{\thetable}{S\arabic{table}}
\setcounter{table}{0} 

\clearpage
\begin{center}
\LARGE\bfseries \section*{Supplementary Materials}
\vspace{2cm}

\LARGE\bfseries Rapid and Automated Alloy Design with Graph Neural Network-Powered LLM-Driven Multi-Agent Systems
\vspace{1cm}

\end{center}
\begin{center}

Alireza Ghafarollahi and Markus J. Buehler

\vspace{1cm}
\noindent \textbf{Correspondence:} \texttt{mbuehler@MIT.EDU}
\end{center}

\renewcommand{\thesection}{S\arabic{section}}

\clearpage

\clearpage
\newpage

\begin{figure}[ht!]
\centering
    \includegraphics[width=.7\textwidth]{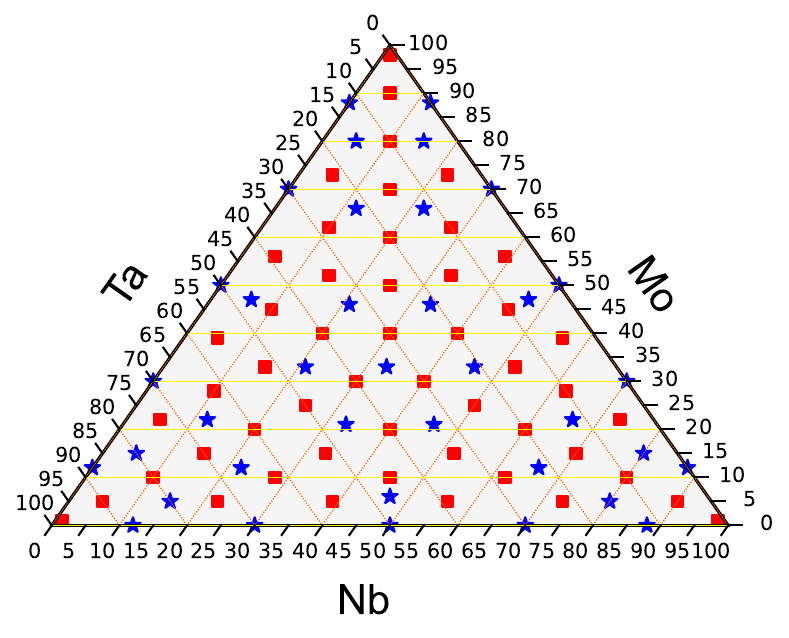}
    \caption{\textbf{Ternary diagram showing compositions for training and validation (red) and new unseen compositions for testing (blue).}}
    \label{fig:ternary_composition}
\end{figure}

\begin{figure}[ht!]
\centering
    \includegraphics[width=1\textwidth]{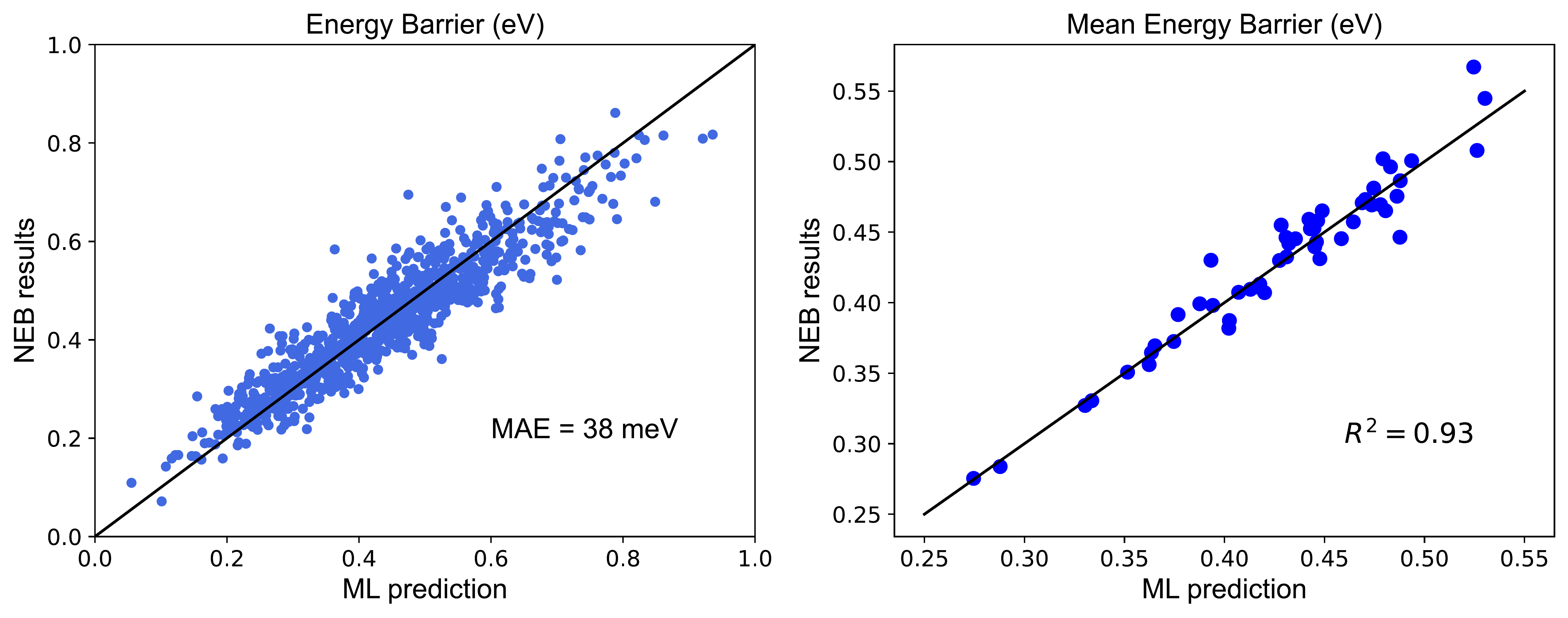}
    \caption{\textbf{Evaluation of the GNN model for potential energy change prediction on the validation set.}}
    \label{fig:energy_barrier_val}
\end{figure}

\begin{figure}[ht!]
\centering
    \includegraphics[width=1\textwidth]{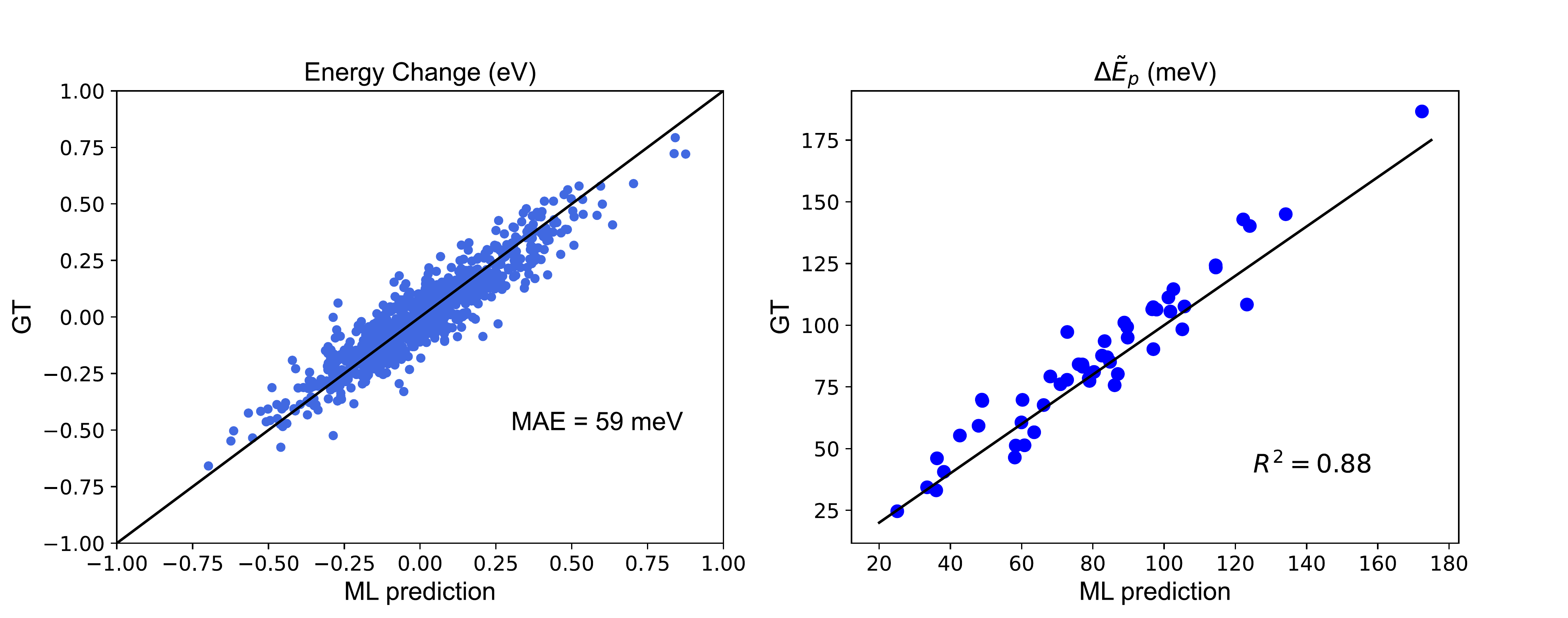}
    \caption{\textbf{Evaluation of the GNN model for Peierls barrier prediction on the validation set.}}
    \label{fig:energy_change_val}
\end{figure}

\begin{figure}[ht!]
\centering
    \includegraphics[width=1\textwidth]{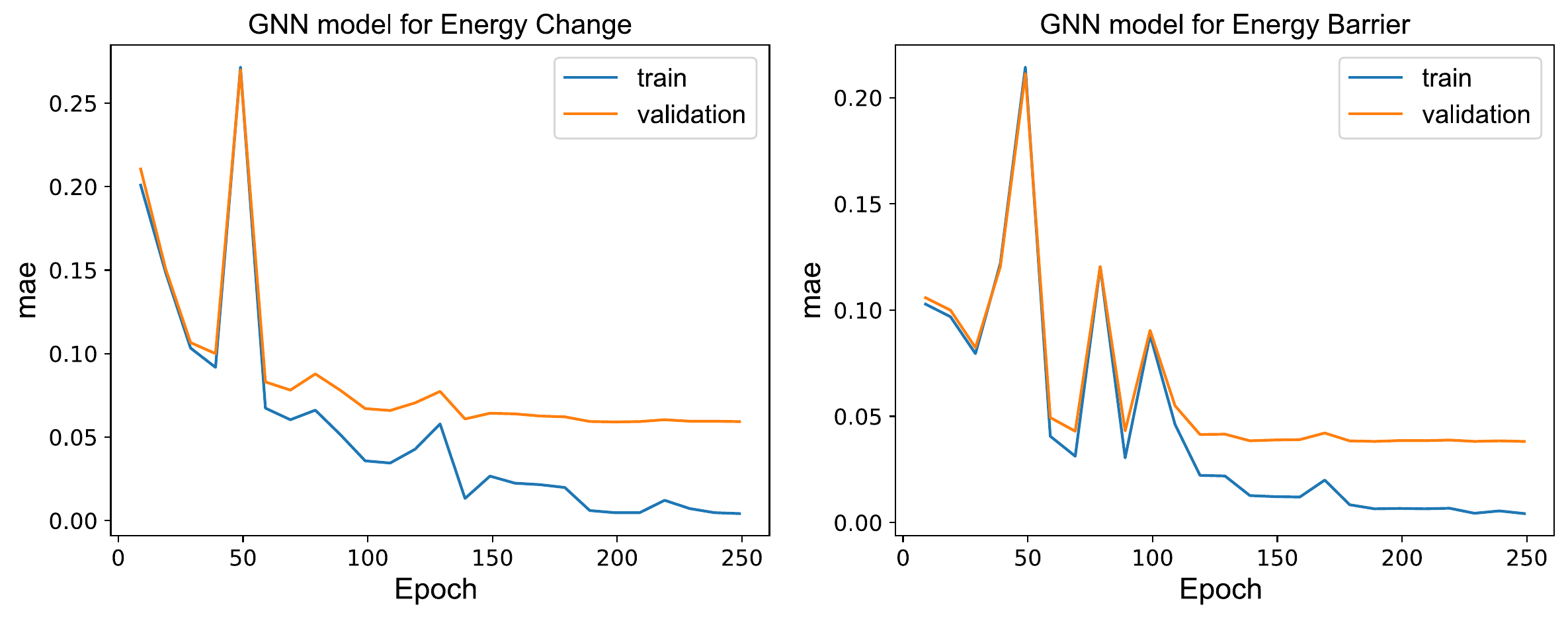}
    \caption{\textbf{Loss evolution during training.} Learning curves over 250 epochs
for the models trained on (left) potential energy changes and (right) Peierls barrier.}
    \label{fig:loss_training_val}
\end{figure}

\begin{figure}[ht!]
    \centering
\begin{Box1}[colbacktitle={white!90!white}, colback={black!3!white}]{Plot analysis (GPT-4o)}
\footnotesize{\texttt{The ternary plot illustrates the variation of the Peierls barrier in the Nb-Mo-Ta alloy system. Here's a detailed analysis:
\\
\\
\#\#\# Trend Identification\\
- **High Peierls Barrier:** The highest values (around 0.095 eV) are observed in compositions with a high percentage of Mo, particularly towards the Mo apex of the triangle. This suggests that Mo significantly contributes to increasing the Peierls barrier.\\
- **Low Peierls Barrier:** The lowest values (around 0.055 eV) are found near the Ta apex, indicating that Ta-rich compositions tend to have lower Peierls barriers.
\\
\#\#\# Comparative Analysis\\
- **Nb Influence:** As the Nb content increases, the Peierls barrier generally decreases, especially when moving from Mo-rich to Nb-rich compositions.\\
- **Mo Influence:** Mo appears to have a strong positive influence on the Peierls barrier, as evidenced by the gradient from the Mo apex towards the other two elements.\\
- **Ta Influence:** Ta tends to lower the Peierls barrier, as seen by the darker colors near the Ta apex.
\\
\\
\#\#\# Statistical Considerations\\
- The gradient of colors suggests a smooth transition in Peierls barrier values across the composition space, indicating consistent trends without abrupt changes.
\\
\\
\#\#\# Contextual Interpretation\\
- **Mechanical Properties:** A higher Peierls barrier typically indicates greater resistance to dislocation motion, suggesting that Mo-rich alloys might exhibit higher strength.\\
- **Applications:** Alloys with higher Mo content could be more suitable for applications requiring high strength, while Ta-rich alloys might be preferred where ductility is more critical.
\\
\\
This analysis provides insights into how varying the composition of Nb, Mo, and Ta can tailor the mechanical properties of the alloy for specific applications.}}
\end{Box1}
    \caption{Results of the analysis performed by the multi-modal agent powered by GPT-4o for the plot shown in Figure \ref{fig:workflow_exp_barrier}(c)}
    \label{fig:plot_analysis_barrier_ternary}
\end{figure}

\begin{figure}[ht!]
    \centering
\begin{Box1}[colbacktitle={white!90!white}, colback={black!3!white}]{Plot analysis (GPT-4o)}
\footnotesize{\texttt{
The plot shows the yield stress versus temperature for various Nb-Mo alloys, comparing the Maresca-Curtin model predictions with experimental data. Here's an analysis of the results:
\\
\\
\#\#\# Trend Identification\\
- **General Trend**: For all alloys, the yield stress decreases with increasing temperature, which is typical for metallic materials.\\
- **Model Predictions**: The Maresca-Curtin model predicts a smooth, nonlinear decrease in yield stress with temperature for each alloy.
\\
\\
\#\#\# Comparative Analysis\\
- **Nb95Mo5**: The model underestimates the yield stress compared to experimental data across the temperature range. The discrepancy increases at lower temperatures.\\
- **Nb91.5Mo8.5**: The model predictions are closer to experimental data, but still slightly underestimate yield stress, especially at lower temperatures.\\
- **Nb81Mo19**: The model aligns well with experimental data, showing only minor discrepancies.\\
- **Nb75Mo25**: The model overestimates the yield stress compared to experimental data, particularly at lower temperatures.
\\
\\
\#\#\# Accuracy of the Maresca-Curtin Model
- The model generally captures the trend of decreasing yield stress with increasing temperature but shows varying accuracy across different alloy compositions.\\
- It performs best for Nb81Mo19, where the predictions closely match experimental data.\\
- The model is less accurate for Nb95Mo5 and Nb75Mo25, where it underestimates and overestimates the yield stress, respectively.
\\
\\
\#\#\# Discrepancies and Possible Reasons\\
- **Alloy Composition**: Variations in alloy composition can affect mechanical properties, and the model may not fully account for these effects.\\
- **Microstructural Factors**: Differences in grain size, phase distribution, or other microstructural features not captured by the model could lead to discrepancies.\\
- **Temperature Effects**: The model may not fully incorporate all temperature-dependent mechanisms affecting yield stress, such as dislocation movement or phase transformations.\\
- **Experimental Variability**: Experimental data can have inherent variability due to measurement errors or sample differences.
\\
\\
\#\#\# Suggestions for Improvement\\
- **Model Refinement**: Incorporating more detailed microstructural and compositional factors could improve model accuracy.\\
- **Parameter Calibration**: Adjusting model parameters based on a broader range of experimental data might enhance predictive capability.\\
- **Further Experiments**: Conducting additional experiments to explore the effects of microstructure and composition could provide insights for model adjustments.
}}
\end{Box1}
    \caption{Results of the analysis conducted by the multi-modal agent powered by GPT-4o for the plot shown in Figure \ref{fig:workflow_exp_strength}(c).}
    \label{fig:plot_analysis_strength_binary}
\end{figure}

\begin{figure}[ht!]
    \centering
\begin{Box1}[colbacktitle={white!90!white}, colback={black!3!white}]{Plot analysis (GPT-4o)}
\footnotesize{\texttt{
The plot shows the yield stress versus temperature for three different Nb-Mo-Ta alloy compositions: Nb33Mo33Ta34, Nb20Mo20Ta60, and Nb20Mo60Ta20. Here's an analysis of the data:
\\
\\
\#\#\# Yield Stress Variation with Temperature
\\
\\
1. **General Trend**:\\
   - For all three alloys, the yield stress decreases as the temperature increases. This is a typical behavior for metals, as higher temperatures generally lead to increased atomic vibrations, reducing the resistance to dislocation motion.
\\
\\
2. **Comparative Analysis**:\\
   - **Nb20Mo20Ta60 (Green Squares)**: This alloy exhibits the highest yield stress across the entire temperature range. It starts at around 950 MPa at low temperatures and decreases to about 200 MPa at 1000 K.\\
   - **Nb33Mo33Ta34 (Blue Circles)**: This alloy has a moderate yield stress, starting at approximately 800 MPa and decreasing to about 150 MPa at 1000 K.\\
   - **Nb20Mo60Ta20 (Red Triangles)**: This alloy shows the lowest yield stress, beginning at around 600 MPa and dropping to about 100 MPa at 1000 K.
\\
\\
\#\#\# Effects of Alloy Composition
\\
\\
- **High Ta Content (Nb20Mo20Ta60)**: The high concentration of Ta (60\%) significantly enhances the yield stress, likely due to Ta's ability to provide solid solution strengthening. Ta atoms create lattice distortions that impede dislocation motion, increasing the yield stress.
\\
\\
- **Balanced Composition (Nb33Mo33Ta34)**: With nearly equal concentrations of Nb, Mo, and Ta, this alloy shows moderate strengthening. The balanced composition may lead to a more uniform distribution of solute atoms, providing moderate lattice distortion and strengthening.
\\
\\
- **High Mo Content (Nb20Mo60Ta20)**: The high Mo concentration results in the lowest yield stress. While Mo can contribute to solid solution strengthening, its effectiveness may be less than that of Ta in this alloy system.
\\
\\
\#\#\# Mechanisms of Solute Strengthening
\\
\\
- **Solid Solution Strengthening**: The primary mechanism at play is likely solid solution strengthening, where the different atomic sizes of Nb, Mo, and Ta create lattice distortions. These distortions hinder dislocation movement, increasing the yield stress.
\\
\\
- **Temperature Effects**: As temperature increases, the effectiveness of solid solution strengthening decreases due to enhanced atomic mobility, which allows dislocations to bypass obstacles more easily.
\\
\\
\#\#\# Insights
\\
\\
- The alloy with the highest Ta content (Nb20Mo20Ta60) shows the greatest resistance to deformation, indicating Ta's significant role in strengthening.\\
- The decrease in yield stress with temperature suggests that these alloys may be more suitable for applications at lower temperatures where higher strength is required.\\
- Understanding the balance of alloying elements can help in designing alloys with desired mechanical properties for specific applications.
}}
\end{Box1}
    \caption{Results of the analysis conducted by the multi-modal agent powered by GPT-4o for the plot shown in Figure \ref{fig:workflow_exp_strength}(d).}
    \label{fig:plot_analysis_strength_ternary}
\end{figure}

\clearpage
\newpage

\begin{figure}[ht!]
    \centering
\begin{Box1}[colbacktitle={white!90!white}, colback={black!3!white}]{}
\footnotesize{\texttt{
You are the Assistant, the central agent in a multi-agent system tasked with solving complex problems. You call different tools and functions, based on a plan, and return the results.
\\
\\
\# REQUIREMENTS\\
- Task Initiation: When given a task, your first step is to use the "plan\_task" tool to generate a plan. If the returned plan includes missing parameters, you should ask the user to provide them.
\\
\\
- Execution: Do not start executing any functions until you receive a plan from the "plan\_task". Follow the plan meticulously to ensure all steps are completed as outlined.
\\
\\
- Error Handling: If an error occurs due to a human mistake, such as the wrong name of a potential, immediately ask the user for the correct information.
\\
\\
- Data Integrity: When calling a function, if you discover that a critical input parameter is missing, prompt the user to provide the necessary data. Do not make assumptions or use your own interpretations to fill in missing information.
\\
\\
\#\# REQUIREMENTS ON Parallel function calling\\
- If the inputs of a function are derived from the results of another function, NEVER execute these functions together, as this becomes problematic. ALWAYS wait for the results of the first function before calling the second one.}}
\end{Box1}
    \caption{Profile of the \textit{Assistant} AI agent.}
    \label{fig:profile_assistant}
\end{figure}

\begin{figure}[ht!]
    \centering
\begin{Box1}[colbacktitle={white!90!white}, colback={black!3!white}]{}
\footnotesize{\texttt{
\#\# Your role
planner. You are a sophisticated planner capable of breaking down complex tasks into simpler subtasks. If the task requires calling or executing functions and tools, clearly specify the function names and their input parameters. These functions should be chosen from the available function library. If the query is in the form of a question/answer that does not require tool calls, you should identify the agent responsible for answering. Your plan will be reviewed by a reviewer for completeness and correctness, so ensure accuracy and attention to detail. After receiving feedback from the critic, revise the plan as necessary.
\\
\\
\#\# Structure Requirements\\
The plan must adhere to the following structure:
\\
\\
- Plan steps: The plan consists of sub-tasks that should be completed sequentially. Specify the key steps in accomplishing the task. This is the overall plan steps, no details are needed here.
\\
\\
- Tools and Input Parameters: If the task requires simulations, coding, plotting, or other visualization analyses, list the required functions and tools. Clearly state the input parameters for each function. If a parameter is unknown or cannot be assumed, note this in the "Unspecified Parameters" section. If no tools are required for the task, return "None" here.
\\
\\
- Unspecified Parameters: List any missing essential input parameters (such as potential name or material name) that should have been provided by the user. If no parameters are missing, state "None."
\\
\\
\#\# Additional Requirements
- Available Tools: Only select tools from the available library. Do not make suggestions or recommendations that involve external tools beyond those available in our domain.
\\
\\
\#\# ONLY IF THE PLAN INVOLVES ALLOY COMPOSITIONS
\\
\\
- Composition Suggestion: Ensure that the plan provides the exact composition(s) in the form of a tuple to be used for the computations or analysis rather than a general response.   
\\
\\
- Composition Interval: For systematic studies involving a range of solute concentrations, the tools "computation\_task\_DEp" and "computation\_task\_PeierlsBarrier" are based on a pre-trained Graph Neural Network. For such studies, if the number of compositions (interval) is not specified, you may use 5\% as the solute interval and construct the compositions. 
\\
\\
- If the task involves function calling for a considerable number of compositions (say>20), you must put all the compositions in a list as input, e.g. $[[10, 10, 80], [10, 80, 10], \ldots [33, 33, 34]]$. 
\\
\\
\#\# REQUIREMENT ON INPUT PARAMETERS\\
- If numeric data are provided as input parameter to the function, the message (or query) should describe the data in great detail.}}
\end{Box1}
    \caption{Profile of the \textit{planner} AI agent.}
    \label{fig:profile_planner}
\end{figure}

\begin{figure}[ht!]
    \centering
\begin{Box1}[colbacktitle={white!90!white}, colback={black!3!white}]{}
\footnotesize{\texttt{
\#\# Your role
You are a sophisticated evaluator. You will be provided with a plan from a planner. Your task is to review the plan suggested by planner. You provide feedback about the plan in an instructive manner. No adjustment is needed if the plan has all the key aspects. Keep your answer short but accurate. 
\\
\\
\#\# REQUIREMENTS
\\
\\
- Ensure the plan details matches the queries in the provided TASK.
}}
\end{Box1}
    \caption{Profile of the \textit{reviewer} AI agent.}
    \label{fig:profile_reviewer}
\end{figure}

\begin{figure}[ht!]
    \centering
\begin{Box1}[colbacktitle={white!90!white}, colback={black!3!white}]{}
\footnotesize{\texttt{
\# Your role\\
You are a sophisticated software developer with expertise in writing python codes. You will write a Python code in a Python block with close attention to the given task.
\\
\\
\# General Requirements\\
- Should the task involve plotting data or saving them in a csv file, the code should be able to save the file in the local computer after execution.
\\
\\
\# Response Instruction\\
- Python code in a python block
\\
\\
\# Other REQUIREMENTS:\\
- Avoid using line plots if there is no inherent ordering (e.g., ascending or descending) in the lables (y-axis, output).
- Avoid using 3d plots and bar plots as they are hard to interpret visually.\\
- Ensure that plot labels, markers, ticks, and titles are large enough for clarity and easy interpretation.\\
- Unless impossible, avoid using subplots and plot the data in a single plot.\\
- Ensure to include the axis labels that clearly describe the parameter. 
\\
\\
\# ERROR HANDLING\\
- if you got "exitcode: 1 (execution failed)", it means the code was not executed successfully. In this case, you SHOULD correct the code. Repeat this until the code is done.
\\
\\
\# WHEN TO TERMINATE\\
- If the code was executed successfully without an error, return 'TERMINATE\_CODER'. NEVER terminate in the first iteration.}}
\end{Box1}
    \caption{Profile of the \textit{coder} AI agent.}
    \label{fig:profile_coder}
\end{figure}

\begin{figure}[ht!]
    \centering
\begin{Box1}[colbacktitle={white!90!white}, colback={black!3!white}]{}
\footnotesize{\texttt{
You are a sophisticated data analyst with deep expertise in identifying trends, patterns, and anomalies in visualized data. Perform a comprehensive analysis of the data presented in the plot or image. 
\\
\\
\#\# As you evaluate, consider the following:
- Trend Identification: Identify overall trends in the data. Are there any linear or nonlinear trends? Are certain regions showing distinct behaviors, such as acceleration or deceleration in the data?\\
- Comparative Analysis: If the plot contains multiple datasets, compare them across relevant dimensions. Are there noticeable differences between groups, or do they follow similar trends? What conclusions can you draw from these comparisons?\\
- Statistical Considerations: Evaluate the plot for any indications of variability or uncertainty. If error bars or confidence intervals are present, assess how they impact the reliability of the conclusions drawn from the data.\\
- Contextual Interpretation: Consider the broader context of the plot. How does the visualized data relate to underlying theories or external factors? Provide insights that can guide further research or decision-making based on the observed trends and behaviors.}}
\end{Box1}
    \caption{Profile of the \textit{multi-modal} AI agent.}
    \label{fig:multi_modal_agent}
\end{figure}

\end{document}